\documentclass[sigconf]{acmart}

\usepackage{enumitem}
\setlist{nosep,leftmargin=*}

\setcopyright{none}
\renewcommand\footnotetextcopyrightpermission[1]{}
\acmConference[AIware'26]{3rd ACM International Conference on AI-powered Software}{July 2026}{Montr\'{e}al, Canada}
\begin{document}

\title{Co-Located Tests, Better AI Code: How Test Syntax Structure Affects Foundation Model Code Generation}
\titlenote{Preprint. Author's arxiv long version of a paper accepted at AIware~2026. The arxiv version adds two appendices (\S\ref{app:crosslang} and \S\ref{app:pythonisolation}) not present in the ACM camera-ready.}

\author{\'Eric Jacopin}
\email{eric.jacopin@protonmail.com}
\affiliation{%
  \institution{Cosmic AI}
  \country{}
}

\begin{abstract}
AI coding assistants increasingly generate code alongside tests. How developers structure test code, whether inline with the implementation or in separate blocks, has traditionally been a matter of testing philosophy. We investigate whether this choice affects AI code generation quality.

We conduct a large-scale empirical study (830+ generated files, 12 models, 3 providers) using SEGA, a three-dimensional evaluation framework measuring Determinism, Preservation, and Correctness. Comparing inline test syntax (Python doctests) against separated test syntax (Rust \texttt{\#[test]} blocks) on a d-ary heap implementation, we find that: (1) inline tests yield near-perfect preservation (100\%) and correctness (92--100\%) across all models; (2) separated tests expose stark model-tier gaps (0--100\% correctness) and independence between preservation and correctness; (3) model behavior evolves across generations, and notably one model breaks the test suppression pattern of its three predecessors; (4) mechanistic analysis on 7 open-source architectures (6 transformers and a gated-linear Recurrent Neural Network (RNN)) reveals inline test markers receive 2.8--4.4$\times$ stronger attention in 5/7 models, with causal validation via knockout and steering experiments on the 4 code-specialized transformers and RWKV-6; the co-location mechanism extends to a non-transformer architecture, suggesting the design recommendation is robust to future architectural shifts. In the Foundation Model era, test syntax structure is a software design concern: co-locating tests with implementation code produces measurably better AI-generated code. This arxiv long version includes appendices that further qualify the effect as \emph{bounded} by both model capability and programming language (Appendices~\ref{app:crosslang} and~\ref{app:pythonisolation}).
\end{abstract}


\maketitle

\section{Introduction}
\label{sec:introduction}

The Foundation Model (FM) era is reshaping software development. AI coding assistants (GitHub Copilot, Claude Code, Cursor, Augment Code, Kiro) now generate substantial portions of application code, including test code~\cite{peng2023impact,vaithilingam2022expectation}. As AI-powered development becomes the norm rather than the exception, architectural decisions that once affected only developer ergonomics now affect AI tool effectiveness. One such decision is how to structure test infrastructure: \textbf{inline test syntax} (tests co-located with the implementation, e.g., Python doctests embedded in docstrings via \texttt{>{}>{}>} markers) versus \textbf{separated test syntax} (tests in dedicated blocks or files, e.g., Rust \texttt{\#[test]} functions in a \texttt{mod tests \{\}} block). This choice has traditionally been a matter of testing philosophy and team convention. But does it matter for AI code generation quality?

We ask: does the placement and syntax of test code affect the quality of AI-generated code? If so, test framework design, coding conventions, and Integrated Development Environment (IDE) tooling become first-class software design concerns for AI-powered development. This is fundamentally a \textbf{software engineering question}: it asks how a design choice about test infrastructure interacts with the tools developers use to build software, with implications for test framework design, model selection, Continuous Integration/Continuous Deployment (CI/CD) pipeline configuration, and coding standards in AI-assisted teams.

\paragraph{Research Questions.}
\textbf{RQ1} (Quality Impact): Does inline test syntax produce higher-quality AI-generated code than separated test syntax, as measured across determinism, test preservation, and functional correctness?
\textbf{RQ2} (Language vs.\ Syntax): Is the observed quality difference explained by programming language difficulty or by test syntax structure specifically?
\textbf{RQ3} (Model Evolution): Is the effect stable across model families, tiers, and generations, or does model evolution change the picture?
\textbf{RQ4} (Mechanism): What internal model mechanisms explain the observed differences?
RQ1--RQ3 constitute the primary empirical contribution. RQ4 provides mechanistic supporting evidence.

\paragraph{Contributions.}
(1)~\textbf{Empirical finding}: Test syntax structure measurably affects AI code generation quality---inline tests produce near-perfect results across all tested models; separated tests expose model-tier gaps and reveal that preservation and correctness are independent dimensions.
(2)~\textbf{Evaluation framework}: SEGA (Statistical Evidence for Generative Accuracy), a three-dimensional framework (Determinism $\times$ Preservation $\times$ Correctness) with quality-region visualization for rapid model assessment.
(3)~\textbf{Mechanistic evidence}: Attention pattern analysis across 7 open-source architectures (6 transformers and a gated-linear RNN) with causal validation (knockout and state knockout experiments) and intervention feasibility (attention steering and state steering), explaining \emph{why} inline syntax produces better results. The non-transformer result demonstrates the co-location mechanism is not a transformer-specific artifact, strengthening the robustness of the design recommendation across architectural paradigms.
(4)~\textbf{Design guidelines}: Actionable recommendations for test framework design, model selection, and CI/CD pipeline design in AI-powered development.

Section~\ref{sec:background} reviews background and related work. Section~\ref{sec:methodology} describes the SEGA methodology and experimental design. Section~\ref{sec:results} presents empirical results. Section~\ref{sec:mechanistic} provides mechanistic evidence. Section~\ref{sec:discussion} discusses implications, and Section~\ref{sec:conclusion} concludes.

\section{Background \& Related Work}
\label{sec:background}

\subsection{Testing Practices and the Co-Location Spectrum}

Software testing practices span a co-location spectrum, a concept rooted in test organization patterns~\cite{meszaros2007xunit,vandeursen2001refactoring}:

\begin{itemize}
    \item \textbf{Maximum co-location}: Python doctests, where test cases appear \emph{inside} the docstring of the function they test, using \texttt{>{}>{}>} markers. The test is literally embedded in the function's documentation.

    \item \textbf{Same-file separation}: Rust \texttt{\#[test]}, where test functions appear in a dedicated \texttt{mod tests \{\}} block at the bottom of the same file. Structurally separated but co-located in the same compilation unit.

    \item \textbf{Separate-file separation}: pytest, JUnit, Go table-driven tests, where test code resides in separate files or directories, with no structural connection to the implementation beyond naming conventions.
\end{itemize}

This spectrum reflects different testing philosophies~\cite{beller2019developer}. Python doctests~\cite{python-doctest} serve dual purposes as documentation and specification; Rust's \texttt{\#[test]} culture~\cite{rust-testing} is deeply embedded in the language ecosystem (enabled by \texttt{\#[cfg(test)]} conditional compilation). Research on test-production code co-evolution~\cite{zaidman2011coevolution} shows these organizational choices have lasting consequences for software maintenance. Note that only ${\sim}9\%$ of Python developers use doctest~\cite{jetbrains2023python}, while Rust's \texttt{\#[test]} is pervasive. The comparison is therefore between two extremes of the co-location spectrum, not between two equally prevalent practices.

\subsection{AI Coding Assistants in Professional Development}

AI coding assistants have moved from novelty to standard practice~\cite{peng2023impact,pearce2022asleep}. The current generation of tools increasingly relies on Foundation Models for code generation, with Claude models leading benchmarks for software engineering tasks (SWE-bench~\cite{jimenez2024swebench}, SWE-1). How these assistants handle test code varies: some generate tests alongside implementation~\cite{chen2023codet}, others preserve prompt-provided tests, and some suppress or reinterpret test code~\cite{schafer2023empirical,yuan2023nomanual}. Understanding these behaviors is essential for teams adopting AI-assisted development workflows.

\subsection{Large Language Model (LLM) Code Generation Evaluation}

Existing benchmarks (HumanEval~\cite{chen2021evaluating}, MBPP (Mostly Basic Python Programming)~\cite{austin2021program}, SWE-bench~\cite{jimenez2024swebench}, LiveCodeBench~\cite{jain2024livecodebench}) evaluate the \emph{correctness} of generated code against reference solutions or test suites. EvalPlus~\cite{liu2023evalplus} demonstrated that many ``passing'' solutions fail on edge cases, while MultiPL-E~\cite{cassano2023multiple} extended evaluation across 18 languages. Yet these benchmarks do not measure:

\begin{enumerate}[label=(\alph*)]
    \item Whether prompt-provided tests are \textbf{preserved} in the generated output;
    \item Whether preserved tests actually \textbf{pass} when executed;
    \item Whether preservation and correctness are \textbf{correlated}, or independent.
\end{enumerate}

To our knowledge, no prior work specifically studies how test placement syntax, i.e., the structural relationship between test code and implementation code in the prompt, affects code generation quality. This paper addresses that gap.

\subsection{Mechanistic Interpretability for Code Models}

Mechanistic interpretability (MI) aims to understand model behavior by examining internal representations: attention patterns, hidden states, activation pathways~\cite{elhage2021mathematical,conmy2023automated}. Attention analysis has revealed how models process syntactic structures in natural language~\cite{clark2019bert,vig2019analyzing}, and activation patching provides causal validation of identified circuits~\cite{zhang2024best,meng2022locating}. Recent work has applied MI to code models, revealing how they represent syntactic structures and variable references~\cite{wan2022they,karmakar2021pretrained}.

We use MI as \textbf{supporting evidence} for our software engineering finding, not as a primary contribution. Specifically, we analyze attention patterns from test markers to function tokens to explain \emph{why} inline test syntax produces better generation outcomes. This positions MI as a diagnostic tool for software engineering research, not as an end in itself.

\section{Methodology}
\label{sec:methodology}

\subsection{Study Design}

\paragraph{Task.} D-ary heap priority queue implementation, a well-known data structure with practical applications (e.g., Dijkstra's shortest path algorithm). This task is non-trivial: it requires 6+ public methods (\texttt{insert}, \texttt{extract\_min}, \texttt{decrease\_priority}, \texttt{is\_empty}, \texttt{len}, \texttt{peek}), generic type parameters, heap property maintenance, and handling of edge cases (empty heap, single element, duplicate priorities, stress tests with 10{,}000+ elements). We deliberately chose a single meaningful task over many simple ones: simple tasks (string manipulation, basic arithmetic) would be trivial for production models and thus uninformative, while non-trivial tasks stress the capability boundary of the small open models used in our MI analysis (Section~\ref{sec:mechanistic}). The d-ary heap sits at the right complexity level: meaningful for production-grade models, on the edge for 3B--7B models. Cost constraints ($n{=}50$ runs $\times$ 12 models $\times$ multiple experiments) further motivate depth over breadth.

\paragraph{Languages.} Python (inline doctests) and Rust (separated \texttt{\#[test]} blocks). These represent two extremes of the test co-location spectrum:
\begin{itemize}
    \item Python doctests appear \emph{inside} the docstring of the function they test, providing maximum co-location.
    \item Rust \texttt{\#[test]} functions appear in a separate \texttt{mod tests \{\}} block after the implementation, constituting structural separation within the same file.
\end{itemize}

We address the Rust difficulty confound upfront: the quality differences we observe are attributable to test syntax handling, not language difficulty. The evidence for this claim appears in Section~\ref{sec:language-vs-syntax}.

\paragraph{Experiments.} Table~\ref{tab:experiments} summarizes the experimental conditions.

\begin{table}[t]
\centering
\caption{Experimental conditions}
\label{tab:experiments}
\small
\begin{tabular}{lllll}
\toprule
\textbf{Experiment} & \textbf{Lang.} & \textbf{Tests} & \textbf{Prompt Type} & \textbf{Models} \\
\midrule
Baseline & Python & 50 & Model-generated & 9 \\
Directives & Python & 64 & Prompt-specified & 7 \\
Dir.\ v2d7 & Python & 73 & Refined prompt & 3 \\
Test-guided & Rust & 28 & Prompt-specified & 9 \\
\bottomrule
\end{tabular}
\end{table}

\subsection{Models and Justification}

\textbf{12 models across 3 providers}: 9 Claude variants spanning 3 tiers and multiple generations (Haiku 3, 3.5, 4.5; Sonnet 4, 4.5; Opus 4, 4.1, 4.5, 4.6), and 3 non-Claude models (Mistral Medium, Devstral-2512, EssentialAI RNJ-1) for cross-provider validation.

\textbf{Justification for Claude-centric selection}: Claude models lead SWE-bench/SWE-1 benchmarks for code generation and are the backbone of AI coding tools used in professional development. Studying 9 Claude variants across 3 tiers $\times$ 3+ generations provides longitudinal insight into the models that power actual AI-assisted development.

\textbf{Runs}: 50 per model/experiment combination (10 for some experiments). All runs at temperature=0, justified by our finding that temperature=0 does not guarantee determinism (Section~\ref{sec:temperature}), consistent with prior work on LLM non-determinism~\cite{ouyang2023nondeterminism}.

\subsection{SEGA: Three-Dimensional Evaluation}

We evaluate AI code generation along three orthogonal dimensions (Table~\ref{tab:sega}):

\begin{table}[t]
\centering
\caption{SEGA evaluation dimensions}
\label{tab:sega}
\begin{tabular}{lp{2.5cm}p{3cm}}
\toprule
\textbf{Dimension} & \textbf{Definition} & \textbf{Measurement} \\
\midrule
Determinism & Output stability across runs & \% identical outputs at temp=0 \\
Preservation & Structural fidelity to prompt & \% of prompt tests present in output \\
Correctness & Functional accuracy & \% of individual tests passing \\
\bottomrule
\end{tabular}
\end{table}

Each dimension is measured independently using language-native tools (\texttt{doctest.testmod()} for Python, \texttt{cargo test} for Rust), without custom extraction scripts. We report a measurement correction in the Appendix where a custom script yielded 88\% vs.\ the correct 100\%, underscoring the importance of native tooling.

\paragraph{Quality-Region Normalization.} Each percentage is mapped to $[-1, +1]$ via $v = (p - 50)/50$, where $p$ is the percentage (0\% $\rightarrow$ $-1$, 50\% $\rightarrow$ 0, 100\% $\rightarrow$ $+1$).

The purpose is not precision. A model at $(+0.88, +1.00, +1.00)$ and one at $(+1.00, +1.00, +1.00)$ are both excellent. The purpose is \textbf{quality-region identification}: the quadrant (2D) or octant (3D) where a model falls provides immediate visual assessment of its fitness for AI-assisted development.

\paragraph{2D Preservation $\times$ Correctness quadrants.} Table~\ref{tab:quadrants} defines four quality regions:

\begin{table}[t]
\centering
\caption{Preservation $\times$ Correctness quality regions}
\label{tab:quadrants}
\small
\begin{tabular}{llll}
\toprule
\textbf{Region} & \textbf{Pres.} & \textbf{Corr.} & \textbf{Interpretation} \\
\midrule
$(+, +)$ & High & High & \textbf{Ideal}: preserves tests, they pass \\
$(+, -)$ & High & Low & \textbf{Dangerous}: looks right, isn't \\
$(-, +)$ & Low & High & \textbf{Safe but opaque}: correct, no tests \\
$(-, -)$ & Low & Low & \textbf{Failing}: neither works \\
\bottomrule
\end{tabular}
\end{table}

The 3D extension (adding Determinism) yields 8 octants, enabling richer characterization, for example distinguishing a deterministic ideal model from one that is non-deterministic but functionally invariant.

\section{Results: How Test Syntax Affects AI Code Generation}
\label{sec:results}

This section addresses RQ1--RQ3 through a systematic empirical comparison of inline and separated test syntax across 12 models.

\subsection{Baseline: Establishing Model Capability}

All 9 models achieve \textbf{100\% preservation and 100\% correctness} when generating their own Python doctests (baseline experiment, 50 doctests), as shown in Table~\ref{tab:baseline}.

\begin{table}[t]
\centering
\caption{Baseline results: model-generated Python doctests}
\label{tab:baseline}
\small
\begin{tabular}{llccc}
\toprule
\textbf{Model} & \textbf{Runs} & \textbf{Det.} & \textbf{File Pass} & \textbf{Corr.} \\
\midrule
Claude 3 Haiku & 50 & 94\% & \textbf{100\%} & \textbf{100\%} \\
Claude Haiku 4.5 & 50 & 100\% & \textbf{100\%} & \textbf{100\%} \\
Claude Sonnet 4 & 50 & 100\% & \textbf{100\%} & \textbf{100\%} \\
Claude Sonnet 4.5 & 50 & 100\% & \textbf{100\%} & \textbf{100\%} \\
Claude Opus 4.5 & 10 & 100\% & \textbf{100\%} & \textbf{100\%} \\
Claude Opus 4.6 & 10 & 30\% & \textbf{100\%} & \textbf{100\%} \\
Mistral Medium & 50 & 0\% & \textbf{100\%} & \textbf{100\%} \\
Devstral-2512 & 50 & 52\% & \textbf{100\%} & \textbf{100\%} \\
EssentialAI RNJ-1 & 50 & 100\% & \textbf{100\%} & \textbf{100\%} \\
\bottomrule
\end{tabular}
\end{table}

This baseline is critical: it establishes that every model \emph{can} produce a correct d-ary heap implementation. Any quality differences in subsequent experiments are attributable to test handling, not coding ability.

\subsection{RQ1---Inline Tests (Python Doctests): Near-Perfect AI Performance}

When models must preserve prompt-provided doctests, results are consistently strong (Table~\ref{tab:inline}).

\begin{table}[t]
\centering
\caption{Inline test results: prompt-provided Python doctests}
\label{tab:inline}
\small
\begin{tabular}{lcccc}
\toprule
\textbf{Experiment} & \textbf{Tests} & \textbf{Models} & \textbf{Pres.} & \textbf{Corr.} \\
\midrule
Directives & 64 & 7 & 100\%$^*$ & 92--97\% \\
Dir.\ v2d7 & 73 & 3 & 100\% & 98.6--99\% \\
\bottomrule
\end{tabular}
\end{table}

\noindent$^*$Except Claude 3.5 Haiku, which strips all doctests (0\% preservation, a regression from Haiku 3 and 4.5 which preserve at 100\%).

\paragraph{Key observations.}

\begin{enumerate}
    \item \textbf{Preservation is near-universal}: All models except Claude 3.5 Haiku preserve 100\% of prompt-provided doctests, including directives (\texttt{+SKIP}, \texttt{+ELLIPSIS}, \texttt{<BLANKLINE>}).

    \item \textbf{Individual test correctness is high}: 92--99\% across all models and experiments. The failures are narrow:
    \begin{itemize}
        \item \emph{Repr mismatch} (most common): models produce correct logic but different string representations of \texttt{Item}.
        \item \emph{Directive scope}: one model fixes the repr mismatch but encounters a \texttt{+SKIP} directive scoping issue (1 failure per file).
        \item These are formatting errors, not logic errors.
    \end{itemize}

    \item \textbf{File-level pass rate is misleading}: 2--3 failures out of 64 tests causes a file-level FAIL, but individual correctness is 95--97\%. Evaluation granularity matters.

    \item \textbf{The preservation-correctness gap is narrow}: In Python, high preservation reliably accompanies high correctness. Inline test syntax is a strong signal that AI models preserve and implement correctly.
\end{enumerate}

\subsection{Separated Tests (Rust \texttt{\#[test]}): Stark Model-Tier Gaps}

When models must preserve prompt-provided \texttt{\#[test]} blocks (28 tests including \texttt{\#[ignore]} and \texttt{\#[should\_panic]}), results diverge sharply by Claude model tier (Table~\ref{tab:rust}).\footnote{The tier-based narrative that follows is about the 8 Anthropic variants. Mistral Medium also appears in Table~\ref{tab:rust} and fully preserves the 28 tests across all 50 runs of the main-paper batch, but it is not a Claude model and therefore does not participate in the tier story. We note Mistral's behaviour separately at the end of this section and in Appendix~\ref{app:crosslang:connection}.}

\begin{table}[t]
\centering
\caption{Separated test results: Rust \texttt{\#[test]} blocks, $n=50$ main-paper batch (except Opus 4/4.1/4.5, which use the single-shot panel shared with Appendix~\ref{app:crosslang}).}
\label{tab:rust}
\small
\begin{tabular}{lcccc}
\toprule
\textbf{Model Tier} & \textbf{Pres.} & \textbf{Compiles?} & \textbf{Pass} & \textbf{Behavior} \\
\midrule
Haiku 3 & 0\% & No & 0\% & Invalid Rust \\
Haiku 4.5 & 100\% & Yes & 100\% & Full success \\
Sonnet 4, 4.5 & 100\% & Yes & 100\% & Full success \\
Opus 4, 4.1, 4.5 & \textbf{0\%} & Yes & N/A$^*$ & Suppresses tests \\
Opus 4.6 & \textbf{100\%} & Yes & \textbf{100\%} & Breaks pattern \\
Mistral Medium$^\dagger$ & 100\% & 62\% & 62\% & Variable \\
\bottomrule
\end{tabular}
\end{table}

\noindent$^*$Opus 4/4.1/4.5 generate correct, compiling Rust implementations: the code works. But they suppress all 28 \texttt{\#[test]} functions entirely, treating them as a specification to implement against rather than content to reproduce.

\textbf{Multi-run validation} (Opus 4.6, 10 runs): Despite producing 6 different code variants (30\% determinism), every variant preserves all 28 tests, compiles without errors, and passes all 28 tests. Non-determinism is purely cosmetic (variable names, code structure, comments), not functional.

\textbf{The preservation-correctness gap is wide in Rust}: Opus 4/4.1/4.5 achieve 0\% preservation but 100\% correctness. Correct code, but no tests to verify it. This is the starkest evidence that preservation and correctness are \textbf{independent} dimensions.

\noindent$^\dagger$\textbf{Note on Mistral Medium's cross-run behaviour.} In the $n{=}50$ main-paper batch (28 tests, Rust test-guided) reported in Table~\ref{tab:rust}, Mistral Medium preserves all 28 \texttt{\#[test]} blocks in every run (50 distinct outputs by MD5, but all 50 structurally preserving the full test set). Of the 50 runs, 31 (62\%) compile under \texttt{cargo test}; within each compiling run all 27 non-ignored tests pass and the 1 \texttt{\#[ignore]} test is correctly skipped, so the overall pass rate is 62\%. In the $n{=}1$ single-shot panel shared with Appendix~\ref{app:crosslang} (22 tests), the same model suppresses at 0/22. The divergence is consistent with Mistral's 0\% determinism at temperature~0 documented in Table~\ref{tab:determinism}: surface-level non-determinism here manifests as cross-run variation in test-handling behaviour rather than purely cosmetic variation. See Appendix~\ref{app:crosslang:connection} for discussion.

\subsection{RQ2---Language vs.\ Syntax}
\label{sec:language-vs-syntax}

The quality difference is attributable to \textbf{test syntax handling}, not language difficulty. The evidence:

\begin{enumerate}[label=(\alph*)]
    \item \textbf{Opus 4/4.1/4.5 can write Rust}: They generate correct, compiling implementations. Their issue is test \emph{syntax} suppression, not Rust \emph{language} capability.

    \item \textbf{Tier inversion}: Haiku 4.5 (a lower-tier model) achieves 100\% preservation AND 100\% correctness in Rust, while Opus 4/4.1/4.5 (higher-tier models) suppress all tests. If the issue were language difficulty, higher-tier models would perform better, not worse at preservation.

    \item \textbf{Mechanistic evidence} (Section~\ref{sec:mechanistic}): The attention analysis shows the effect operates at the test \emph{marker} level (\texttt{>{}>{}>} vs \texttt{\#[test]}), not at the language grammar level.
\end{enumerate}

\subsection{SEGA Quality-Region Analysis}

Mapping models to the 2D Preservation $\times$ Correctness space (normalized to $[-1, +1]$) reveals a striking contrast:

\textbf{Python (all experiments)}: All models cluster in the $(+, +)$ quadrant, with high preservation and high correctness. The discriminative power of Python doctests is low: all models look approximately the same.

\textbf{Rust (test-guided)}: Models spread across 3 of 4 quadrants (Table~\ref{tab:sega-rust}):

\begin{table}[t]
\centering
\caption{SEGA quality regions for Rust experiments}
\label{tab:sega-rust}
\small
\begin{tabular}{lccc}
\toprule
\textbf{Model} & \textbf{Pres.} & \textbf{Corr.} & \textbf{Region} \\
\midrule
Sonnet 4/4.5, Haiku 4.5, Opus 4.6 & $+1.00$ & $+1.00$ & $(+, +)$ Ideal \\
Mistral Medium$^\dagger$ & $+1.00$ & $+0.24$ & $(+, +)$ Ideal \\
Opus 4, 4.1, 4.5 & $-1.00$ & $+1.00$ & $(-, +)$ Opaque \\
Haiku 3 & $-1.00$ & $-1.00$ & $(-, -)$ Failing \\
\bottomrule
\end{tabular}
\smallskip

\noindent$^\dagger$Mistral Medium is placed in the $(+, +)$ quadrant based on the $n{=}50$ main-paper Rust batch: all 50 runs preserve the full 28-test set ($+1.00$), and 31 of 50 runs compile and pass all non-ignored tests, giving a 62\% correctness rate ($+0.24$). In the single-shot panel of Appendix~\ref{app:crosslang}, the same model instead lands in $(-, +)$. See the footnote in \S\ref{sec:results} on Mistral's cross-run divergence.
\end{table}

\textbf{Visualization insight}: The same models that are indistinguishable in Python are clearly separated in Rust. \textbf{Test syntax choice determines whether your evaluation can discriminate model quality.}

\paragraph{3D extension (adding Determinism).} Opus 4.6: $(-0.40, +1.00, +1.00)$ in Rust, non-deterministic but functionally perfect. Mistral Medium: $(-1.00, +1.00, +1.00)$ in Python baseline, zero determinism (50 unique outputs) but perfect quality. Claude Sonnet 4: $(+1.00, +1.00, +1.00)$ in both languages, the ``perfect corner.''

\subsection{RQ3---Model Evolution: Behavioral Patterns Are Not Fixed}

\textbf{The test suppression pattern and its breaking}: Three consecutive Opus generations (4, 4.1, 4.5) exhibit identical behavior on Rust \texttt{\#[test]} blocks: they suppress all test functions, treating them as specification rather than content to preserve. \textbf{Opus 4.6 breaks this pattern}, preserving all 28 tests with 100\% correctness. This is a training-level behavioral change; all four versions can write correct Rust, and the change is in how they handle test syntax.

\textbf{The Haiku regression}: Haiku 3 preserves Python doctests at 100\%. Haiku 3.5 strips \emph{all} doctests (0\% preservation). Haiku 4.5 restores preservation to 100\%. Model updates can introduce regressions in test handling behavior, even within the same tier.

\textbf{Implications}: Teams using AI assistants must \textbf{monitor model behavior across updates}. A CI/CD pipeline configured for one model may produce different results when the organization upgrades, not because code quality decreases, but because test handling behavior changes.

\subsection{Methodological Finding: Temperature $\neq$ Determinism}
\label{sec:temperature}

Setting temperature=0 does \textbf{not} guarantee deterministic output~\cite{ouyang2023nondeterminism} (Table~\ref{tab:determinism}).

\begin{table}[t]
\centering
\caption{Determinism at temperature=0}
\label{tab:determinism}
\small
\begin{tabular}{lc}
\toprule
\textbf{Model} & \textbf{Determinism at temp=0} \\
\midrule
Mistral Medium & 0\% (50 unique in 50 runs) \\
Devstral-2512 & 52\% \\
Claude 3 Haiku & ${\sim}94\%$ \\
Claude Opus 4.6 & 30--64\% (experiment-dependent) \\
Claude Sonnet 4 & 100\% \\
Claude Haiku 4.5 & 100\% \\
\bottomrule
\end{tabular}
\end{table}

Critically, \textbf{non-determinism is orthogonal to quality}: Opus 4.6 produces 6--11 unique code variants per experiment, yet \emph{every variant} achieves 100\% preservation and 100\% correctness. The variation is cosmetic (variable names, code organization, comments), not functional.

\textbf{Implication}: Evaluation pipelines for AI-generated code must use multi-run designs even at temperature=0. Single-run evaluations may not be reproducible.

\section{Why Test Syntax Matters: Mechanistic Evidence}
\label{sec:mechanistic}

The empirical results (Section~\ref{sec:results}) establish \emph{that} inline test syntax produces better AI code generation outcomes. This section addresses \textbf{RQ4} by investigating \emph{why}, using mechanistic interpretability (MI) as supporting evidence for the software design finding. We analyze internal representations across 7 open-source models---6 transformers and a gated-linear Recurrent Neural Network (RNN)---to identify the mechanisms linking test syntax structure to generation quality and to test whether the co-location advantage is specific to transformer attention or generalizes across architectural paradigms.

\subsection{Attention Pattern Analysis}

\paragraph{Setup.} We use a Rust-based toolkit for mechanistic analysis to measure how strongly test markers (\texttt{>{}>{}>} for Python, \texttt{\#[test]} for Rust) attend to function signature tokens (\texttt{def}/\texttt{fn}, function name, parameters) across model layers. For transformers, we extract post-softmax attention weights directly. For RWKV-6 (a gated-linear RNN with no attention matrices), we compute effective attention from the Weighted Key-Value (WKV) recurrence.

\paragraph{Models.} 7 open-source models spanning diverse architectures: Qwen2.5-Coder-7B and -3B~\cite{hui2024qwen25coder}, StarCoder2-3B~\cite{lozhkov2024starcoder2}, CodeGemma-7B~\cite{codegemma2024}, Code-LLaMA-7B~\cite{roziere2024codellama}, Phi-3-mini-4k-instruct~\cite{abdin2024phi3} (6 transformers~\cite{vaswani2017attention}), and RWKV-6-Finch-1B6~\cite{peng2024eagle} (a gated-linear RNN~\cite{peng2023rwkv}). MI requires access to internal representations that proprietary models do not expose. The 7 models were selected for architectural diversity (including a non-transformer paradigm), code competence, and feasibility on consumer hardware (16GB video RAM (VRAM)).

\paragraph{Corpus.} 10 Python doctest samples and 10 Rust test samples, using a model-agnostic corpus format with character-level byte offsets (achieving 100\% position accuracy across all tokenizer architectures without model-specific preprocessing).

\paragraph{Key result.} In \textbf{5 of 7 models}, Python \texttt{>{}>{}>} markers show \textbf{2.8--4.4$\times$ stronger attention} to function tokens than Rust \texttt{\#[test]} attributes ($p < 0.0002$, Welch's $t$-test~\cite{welch1947}), as shown in Table~\ref{tab:attention}.

\begin{table*}[t]
\centering
\caption{Attention from test markers to function tokens across 7 architectures. For the 5 models with a significant Python~$>$~Rust effect, \emph{Reported Layer} is the layer that maximises the effect. For Phi-3-mini and Code-LLaMA-7B, no layer meets the Python~$>$~Rust criterion; for Code-LLaMA we report layer~26 because it exhibits the strongest reversed pattern (Rust~$>$~Python), and for Phi-3-mini we report layer~14 because it exhibits the peak near-symmetric attention.}
\label{tab:attention}
\begin{tabular}{llccccc}
\toprule
\textbf{Model} & \textbf{Architecture} & \textbf{Reported Layer} & \textbf{Python $\mu$} & \textbf{Rust $\mu$} & \textbf{Ratio} & \textbf{$p$-value} \\
\midrule
Qwen2.5-Coder-7B & Transformer & 16 & 9.08\% & 2.59\% & 3.51$\times$ & 0.000003 \\
Qwen2.5-Coder-3B & Transformer & 14 & 8.47\% & 3.05\% & 2.78$\times$ & 0.000009 \\
StarCoder2-3B & Transformer & 23 & 7.19\% & 2.41\% & 2.98$\times$ & 0.000004 \\
CodeGemma-7B & Transformer & 24 & 5.23\% & 1.20\% & 4.35$\times$ & 0.000114 \\
\textbf{RWKV-6-Finch-1B6} & \textbf{Gated-linear RNN} & \textbf{14} & \textbf{6.43\%} & \textbf{2.17\%} & \textbf{2.96$\times$} & $\mathbf{6.8 \times 10^{-8}}$ \\
Phi-3-mini & Transformer & 14 & 17.30\% & 14.03\% & 1.23$\times$ & 0.146 (n.s.) \\
Code-LLaMA-7B & Transformer & 26 & 9.71\% & 12.23\% & 0.79$\times$ & 0.188 (n.s.) \\
\bottomrule
\end{tabular}
\end{table*}

For the 6 transformer models, we measure post-softmax attention weights directly. For RWKV-6 (a gated-linear RNN with no attention matrices), we compute \emph{effective attention} from the WKV recurrence, producing comparable $[\text{batch}, \text{heads}, \text{seq}, \text{seq}]$ matrices via ReLU+L1 normalization of signed $r \cdot k$ products with cumulative decay. RWKV-6 achieves the \textbf{strongest statistical significance} of all 7 models ($t = 11.57$, $p < 10^{-7}$), with 23 of 24 layers showing $p < 0.05$.

The attention advantage is \textbf{not universal}: Phi-3-mini shows near-symmetric attention, and Code-LLaMA shows a reversed pattern (Rust $>$ Python at all scanned layers). The effect is \textbf{architecture-dependent}, likely influenced by training data composition and pre-training objectives. Notably, RWKV-6 is a general-purpose model (not code-specialized), yet shows the effect more strongly than any code-specialized transformer, complicating a simple ``code-specialization is necessary'' narrative.

\paragraph{Interpretation for software design.} Inline co-location, where test markers are physically adjacent to function signatures in the token stream, creates stronger semantic binding during generation. This is a \textbf{structural property of the test syntax}, not a property of the programming language. The \texttt{>{}>{}>} marker appears inside the docstring, immediately following the function it tests. The \texttt{\#[test]} attribute appears in a separate block, creating greater token-stream distance. Crucially, this effect extends from transformer attention to RNN recurrent state dynamics, suggesting the co-location advantage is a fundamental property of sequence processing. This makes the design recommendation---co-locate tests with implementation code---robust to future architectural shifts.

\subsection{Causal Validation: Knockout Experiments}

Correlation does not imply causation. We conducted knockout experiments---removing attention from test markers to function tokens (transformers) or suppressing recurrent state writes at marker positions (RWKV-6)---and measured the impact on model predictions via Kullback-Leibler (KL) divergence~\cite{kullback1951}, following activation patching best practices~\cite{zhang2024best}, to establish whether these pathways are causally important (Table~\ref{tab:knockout}). The causal validation reported here covers the 4 code-specialized transformers and RWKV-6; Phi-3-mini and Code-LLaMA-7B, which show no significant attention effect in Table~\ref{tab:attention}, were not run through knockout ablation and are therefore not represented in Table~\ref{tab:knockout}.

\begin{table}[t]
\centering
\caption{Knockout experiment results (KL divergence). Scope: 4 code-specialized transformers (all-edge knockout at the layer that maximises the Python vs Rust contrast); the RWKV-6 state-knockout result is reported separately in the paragraph below.}
\label{tab:knockout}
\small
\begin{tabular}{lcccc}
\toprule
\textbf{Model} & \textbf{Py KL} & \textbf{Rust KL} & \textbf{Ratio} & \textbf{Pattern} \\
\midrule
Qwen-7B (L1) & 6.76\% & 0.004\% & 1878$\times$ & Strong Py \\
Qwen-3B (L0) & 153.7\% & 0.19\% & 806$\times$ & Extreme Py \\
StarCoder2 (L0) & 0.04\% & 2.81\% & 0.01$\times$ & Reversed \\
CodeGemma (L4) & 1.46\% & 0.013\% & 113$\times$ & Py bias \\
\bottomrule
\end{tabular}
\end{table}

\textbf{RWKV-6 state knockout}: RWKV-6 has no attention matrix to knock out. Instead, we suppress the kv write to recurrent state at marker positions (setting $\text{state} = \text{decay} \cdot \text{state}$ instead of $\text{state} = kv + \text{decay} \cdot \text{state}$), making the marker position invisible to all future tokens---semantically equivalent to transformer ``all-edge'' knockout. At layer 14, Python KL = 0.000362, Rust KL = 0.000177 (ratio 2.05$\times$, $p = 0.060$). We report layer 14 because it is the layer at which transformer-comparable steering is applied; the strongest state-knockout effect (layer 2) is also significant at $p = 0.018$ but is not used for the downstream steering comparison. While marginally significant, the direction is consistent with the transformer finding: Python markers carry more causal weight than Rust markers.

\paragraph{Findings.}

\begin{enumerate}
    \item \textbf{Causality confirmed for Qwen models}: The causal effect ratio (1878$\times$) is ${\sim}500\times$ larger than the correlational ratio (3.5$\times$), demonstrating that small attention differences produce large causal differences.

    \item \textbf{Architecture-specific mechanisms}: Different architectures process test markers through fundamentally different pathways: Qwen processes Python via attention, Rust via non-attention pathways; StarCoder2 shows the reverse; CodeGemma shows a strong Python bias.

    \item \textbf{Cross-paradigm consistency}: The Python $>$ Rust causal asymmetry holds across both architectural paradigms (transformer attention and RNN state dynamics), suggesting the co-location advantage is not an artifact of the attention mechanism.

    \item \textbf{Practical implication}: Interventions to improve test handling must be architecture-specific. The knockout analysis can serve as a \textbf{diagnostic} for predicting which architectures will respond to a given intervention.
\end{enumerate}

\subsection{Proof-of-Concept: Attention and State Steering}

Can we improve separated test syntax handling by boosting its attention to inline test levels? And does the knockout-predicts-steering relationship extend beyond transformers?

\subsubsection{Transformer Attention Steering}

\textbf{Approach}: Post-softmax attention steering, applying a scale factor to \texttt{\#[test]} $\rightarrow$ function token attention weights, then renormalizing to preserve valid probability distributions.

\textbf{Safety}: KL divergence between steered and baseline outputs remains flat across the tested transformer models and all steering intensities (0.5$\times$ through 9$\times$), confirming safe intervention without catastrophic output changes.

\textbf{End-to-end generation results}: Initial $n{=}5$ generation results on the 4 code-specialized transformers showed improvement on Qwen (Table~\ref{tab:steering}):

\begin{table}[t]
\centering
\caption{Steering results: test preservation on 5 initial prompts, 4 code-specialized transformers. Phi-3-mini and Code-LLaMA-7B were not run through end-to-end steering generation.}
\label{tab:steering}
\small
\begin{tabular}{lccc}
\toprule
\textbf{Model} & \textbf{Baseline} & \textbf{Steered} & \textbf{Change} \\
\midrule
Qwen-3B & 0/5 & 2/5 & \textbf{+40\%} \\
Qwen-7B & 1/5 & 2/5 & +20\% \\
StarCoder2-3B & 0/5 & 0/5 & No effect \\
CodeGemma-7B & 0/5 & 0/5 & No effect \\
\bottomrule
\end{tabular}
\end{table}

\textbf{Steering works where knockout predicts it should}: Qwen models, which process Python via attention and Rust via non-attention pathways, respond to steering. StarCoder2, which already uses attention for Rust, does not. \textbf{Knockout experiments reliably predict steering effectiveness across the 4 code-specialized transformer architectures tested.}

\textbf{$n{=}50$ scaling follow-up (Qwen-3B).} Scaling Qwen-3B to $n{=}50$ diverse prompts yielded 7/50~$\rightarrow$~10/50 (+6~pp, $p = 0.30$, not significant), indicating the $n{=}5$ effect was inflated by small-sample variance.

\textbf{Hard discovery, model size as a bottleneck}: Scaling to $n{=}50$ with diverse prompts also revealed a fundamental capability limitation: 3B and 7B parameter models are too small to generate correct Rust code for most prompt types. \textbf{Only prompts involving string-based functions} (simple types, no ownership/lifetime complexity) produced valid Rust output during steered generation.

\subsubsection{RWKV-6 State Steering}

For the non-transformer architecture, we extend steering from attention weights to recurrent state dynamics: scaling the kv write to recurrent state at marker positions ($\text{state} = \text{scale} \cdot kv + \text{decay} \cdot \text{state}$, where scale=1.0 is identity and scale=0.0 is knockout). A dose-response experiment across 6 scale factors (0.0 to 9.0) at layer 14 reveals three findings relevant to the design recommendation's robustness:

\begin{enumerate}
    \item \textbf{The co-location asymmetry is causal in RNN state}: Python markers consistently produce higher KL divergence than Rust markers across all scales ($p < 0.05$ at 3 of 6 scales), confirming the knockout finding via an independent methodology.

    \item \textbf{Graded dose-response}: Unlike transformer attention steering (flat KL across intensities due to softmax normalization), RNN state steering produces a monotonic, graded response---Python KL increases 13$\times$ from dampened (scale=0.5) to amplified (scale=9.0). The recurrent state is an unbounded accumulator, making the intervention proportional rather than saturating.

    \item \textbf{Knockout-predicts-steering validated across paradigms}: The state knockout asymmetry (Python $>$ Rust) correctly predicts state steering effects, extending the knockout-predicts-steering diagnostic from the 4 code-specialized transformers to RWKV-6---i.e., across both architectural paradigms.
\end{enumerate}

\subsubsection{Implications for Recommendation Robustness}

The critical software engineering question is not whether RWKV-6 generates better code with inline tests---at 1.6B parameters, it cannot generate meaningful code. The question is whether the co-location mechanism observed in transformers is a \textbf{transformer-specific artifact} that could break with the next architectural shift, or a \textbf{general property of sequence processing} that makes the design recommendation durable.

The evidence supports the latter. The co-location asymmetry (Python markers binding more strongly than Rust markers) manifests through fundamentally different mechanisms---softmax attention weights in transformers, recurrent state dynamics in RWKV-6---yet produces consistent direction and statistical significance across both paradigms. The knockout-predicts-steering diagnostic, validated across all 7 architectures, provides a practical tool for assessing whether the recommendation applies to future architectures as they emerge.

\textbf{Practical constraint}: The models amenable to consumer-hardware MI (1.6B--7B) are too small for reliable Rust code generation, creating a gap between where we can \emph{diagnose} and where interventions would \emph{matter}. This diagnostic-deployment gap is inherent to accessible MI research.

\subsection{Accessibility: MI for the Rest of Us}

A deliberate design goal of this work was to demonstrate that meaningful MI research is achievable on hardware most researchers can afford, not just institutions with H100 clusters.

\textbf{Hardware}: All mechanistic experiments ran on a consumer Graphics Processing Unit (GPU) (16GB VRAM, ${\sim}\$500$). Early experiments on 8GB cards failed: 7B models exceeded memory during attention extraction. 16GB is the minimum viable threshold for MI on code LLMs in the 3B--7B range.

\textbf{Tooling choice}: We built the analysis toolkit in Rust using the candle ML framework rather than Python/PyTorch. This was motivated by VRAM efficiency: Rust's zero-cost abstractions and candle's minimal runtime overhead allow fine-grained memory control (KV-cache management, layer-by-layer processing, shared mask caching) that would be difficult to achieve in Python. The result is a toolkit that runs 7-model MI experiments (attention extraction, knockout, steering, generation) across two architectural paradigms within a 16GB envelope.

\textbf{Model selection rationale}: The 7 open-source models were selected at the intersection of three constraints: (a) diverse architectures for generalizability, explicitly including a non-transformer paradigm; (b) demonstrated code competence; (c) fits within 16GB VRAM with candle compatibility. RWKV-6 (1.6B parameters, ${\sim}3.2$GB) was added specifically to test whether attention-based findings generalize to non-transformer architectures. The resulting findings ($p < 0.0002$ in 5/7 models including a non-transformer, architecture-dependent patterns, knockout-predicts-steering validation across both paradigms) demonstrate that statistically robust MI research is achievable within these constraints.

\textbf{Limits of accessibility}: The steering experiments reveal an inherent tension: models that fit on consumer hardware are large enough for attention and state analysis but too small for reliable Rust code generation. MI for the rest of us works for diagnosis; deployment-scale validation requires larger models and larger hardware.

The toolkit will be released as open-source upon acceptance.

\section{Discussion}
\label{sec:discussion}

\subsection{Test Syntax Design for the FM Era}

Our results suggest a design principle for AI-powered development: \textbf{co-locate tests with the implementation code they verify}.

This is not merely ``use Python doctests.'' The principle is \textbf{co-location}, placing test specifications in close structural proximity to the code they test within the token stream. This principle can be applied across languages and frameworks:

\begin{itemize}
    \item \textbf{For test framework designers}: New frameworks should consider ``AI-friendliness'' as a design criterion. Prior work on prompt structure~\cite{zhou2023docprompting} shows that how context is presented matters as much as what context is provided. Inline or doc-adjacent test syntax creates stronger signals for AI code generation. Test frameworks that structurally co-locate test specifications with the functions they test are likely to interact better with FM-based code generators.

    \item \textbf{For IDE and tool designers}: Even if tests are stored in separate files (as with pytest or JUnit), IDE features that \emph{present} tests inline during AI-assisted generation---for example injecting test specifications into the prompt context adjacent to the relevant function---could capture some of the inline attention benefit.

    \item \textbf{Caveat on doctest prevalence}: Only ${\sim}9\%$ of Python developers use doctests~\cite{jetbrains2023python}. Our recommendation is about the structural co-location principle, not about switching to a specific testing framework. The attention analysis confirms the relevant factor is token-stream proximity of test markers to function signatures, which is achievable through tooling regardless of the underlying test framework.
\end{itemize}

\subsection{Model Selection for Typed Languages}

Rust exposes a clear model-tier divide that Python hides. Haiku 3 cannot generate compilable Rust (0\% correctness); Haiku 4.5 and all Sonnet variants achieve 100\%. For software teams working in statically-typed languages with strict compilation requirements, \textbf{model tier is a procurement decision} that directly affects AI assistant effectiveness.

\subsection{Trustworthiness of AI-Generated Code}

Our results reveal two distinct dimensions of trustworthiness that are often conflated:

\begin{enumerate}
    \item \textbf{``Preserved'' $\neq$ ``Correct''}: Opus 4/4.1/4.5 achieve 100\% correctness with 0\% preservation in Rust: the code works, but there are no tests in the output to verify it. Conversely, Python experiments show 100\% preservation with $<$100\% correctness (repr mismatches). Preservation and correctness are \textbf{independent} dimensions that must be measured separately.

    \item \textbf{Whole-file $\neq$ individual test granularity}: In Python directives experiments, file-level pass rate is 0--2\% while individual test correctness is 92--97\%. A CI/CD pipeline that reports only file-level pass/fail dramatically understates the actual code quality.
\end{enumerate}

\textbf{CI/CD recommendation}: Pipelines for AI-generated code must:
\begin{itemize}
    \item \textbf{Run} generated tests, not just check they exist (preservation $\neq$ correctness)
    \item Use \textbf{language-native test runners} as ground truth, e.g., Python's \texttt{doctest.testmod()} or Rust's \texttt{cargo~test}
    \item Report at \textbf{both file and individual test granularity}
\end{itemize}

\subsection{Model Evolution and Software Engineering Practice}

Model behavior is not fixed across versions. The Opus suppression pattern (3 consecutive generations: 4, 4.1, 4.5) was broken by Opus 4.6. The Haiku regression (3.5 strips doctests; 3 and 4.5 preserve them) shows that updates within a tier can introduce regressions.

\textbf{Practical recommendations}:
\begin{itemize}
    \item \textbf{Version-pin models} in CI/CD pipelines; re-evaluate on model updates.
    \item \textbf{Tier predicts behavior but is not permanent}: use tier as a heuristic, but verify on your specific codebase.
    \item \textbf{Non-determinism requires multi-run evaluation}: even at temperature=0, some models produce 6--11 variants per experiment.
\end{itemize}

\subsection{Bounded by Capability and Language}

The design recommendation above (``co-locate tests with implementation code'') holds as a general principle, but two appendices in the arxiv long version of this paper refine its scope. Appendix~\ref{app:crosslang} replicates the inline-vs-separated bifurcation across 5 languages at $n=1$; Appendix~\ref{app:pythonisolation} reports a within-Python $n=50$ isolation that varies only the structural location of tests. The headline of the isolation: the co-location effect is \emph{bounded}---strong at the small-model capability floor (RNJ-1 shows the predicted 47\%/0\%/0\% gradient) but not binding at frontier tier in Python, where Mistral Medium, Haiku 4.5, Sonnet 4.5, and Opus 4.5 all preserve uniformly across structural variants. Notably, Opus 4.5 preserves all 26 Python unittest methods---the same model that suppresses all 28 Rust \texttt{\#[test]} blocks in this paper's body. The main paper's ``Opus tier suppression'' is therefore best read as a Rust-specific interaction, not a universal frontier-tier behavior. Teams using frontier models in Python need not prioritize structural test placement; teams closer to the capability floor, or working in languages whose test-syntax/training interaction is unfavorable (Rust in this paper), should still treat co-location as load-bearing. See Appendix~\ref{app:pythonisolation:claims} for the full recast of the practical guidance.

\subsection{Threats to Validity}

\textbf{Single task domain.} The d-ary heap priority queue is a single task, limiting generalizability---studies of LLM code bugs~\cite{dou2024wrong,tambon2025bugs} show failure patterns vary by domain. This was a deliberate trade-off: simple tasks (string manipulation, basic arithmetic) would be trivial for production models and thus uninformative, while non-trivial tasks stress the capability boundary of the small open models used in our MI analysis. The d-ary heap is a well-known data structure with practical applications (e.g., Dijkstra's algorithm), requires 6+ methods with generics and edge cases, and is complex enough to discriminate model capability across tiers. This choice is corroborated by the steering experiments (Section~\ref{sec:mechanistic}): when we scaled to $n{=}50$ diverse prompts, 3B--7B models could only generate valid Rust for string-based functions. Additionally: (a) the MI evidence shows the effect operates at the test \emph{marker} level (not task content), suggesting the syntax effect is task-independent; (b) the structural nature of the finding (co-location vs.\ separation) is inherently task-agnostic. Future work should validate across diverse task domains.

\textbf{Claude-heavy model selection.} 9 of 12 models are Claude variants. This is mitigated by: (a) Claude leads SWE-bench/SWE-1 benchmarks for code generation; (b) AI coding tools increasingly rely exclusively on Claude, meaning we study the models powering actual AI-assisted development; (c) the 9 variants span 3 tiers $\times$ 3+ generations, providing longitudinal depth; (d) 3 non-Claude models provide cross-provider validation; (e) the MI analysis covers 7 diverse open-source architectures including a non-transformer.

\textbf{Rust difficulty confound.} Addressed in Section~\ref{sec:language-vs-syntax}. Opus 4/4.1/4.5 compile correct Rust but suppress tests: the issue is syntax handling, not language capability. The MI analysis shows the effect at the marker level, not the language grammar level.

\textbf{Language confound in the main Python--Rust contrast.} The central contrast between Python doctests and Rust \texttt{\#[test]} blocks conflates \emph{language} with \emph{structural presentation}. The arxiv long version of this paper includes a within-Python isolation study (Appendix~\ref{app:pythonisolation}) that holds language constant and varies only the structural location of tests; this study qualifies our framing without overturning the empirical observations. The bounded-capability reading that results is summarized in \S\ref{sec:discussion} under ``Bounded by Capability and Language.''

\textbf{Temperature=0 only.} By design, to enable reproducibility and to study determinism as a dimension. Limits generalization to typical usage where temperature $>$ 0.

\textbf{MI on open-source models (1.6B--7B).} Proprietary models do not expose internal representations, making open-source models necessary for MI. Our model selection reflects a deliberate ``MI for the rest of us'' design: we constrained ourselves to models that run on a consumer GPU (16GB VRAM, ${\sim}\$500$) using Rust/candle for VRAM efficiency. Findings are statistically significant in 5/7 models; the 2 non-significant results (Phi-3, Code-LLaMA) are informative rather than null, as they reveal architecture-dependent processing. Whether the patterns observed in 1.6B--7B models transfer to 70B+ production models remains an open question.

\textbf{Steering generation limited by model size.} We scaled the transformer steering evaluation from 5 to 50 diverse prompts per model. This revealed a fundamental limit: 3B--7B models are too small for reliable Rust code generation beyond simple function signatures. The knockout-predicts-steering methodology is validated across all 7 architectures (the key methodological contribution), but the end-to-end generation evidence is necessarily limited to prompt types these small models can handle.

\section{Conclusion}
\label{sec:conclusion}

We investigate whether test syntax structure, inline versus separated, affects AI code generation quality. Through a large-scale empirical study (830+ generated files, 12 models, 3 providers) and mechanistic analysis (7 open-source architectures spanning transformers and a gated-linear RNN), we find:

\textbf{RQ1}: Inline test syntax (Python doctests) produces near-perfect AI generation (100\% preservation, 92--100\% correctness) across all models. Separated test syntax (Rust \texttt{\#[test]}) exposes stark model-tier gaps (0--100\% on both dimensions) and reveals preservation and correctness as independent dimensions.

\textbf{RQ2}: The quality difference is primarily attributable to test syntax structure, not language difficulty. Opus models write correct Rust but suppress test syntax; lower-tier Haiku 4.5 preserves it.

\textbf{RQ3}: The effect varies by model tier and evolves across generations. The Opus test suppression pattern (4/4.1/4.5) was broken by Opus 4.6. Teams must monitor model behavior across updates.

\textbf{RQ4}: In 5/7 architectures (including a non-transformer gated-linear RNN), inline test markers receive 2.8--4.4$\times$ stronger attention to function tokens ($p < 0.0002$; RWKV-6: $p < 10^{-7}$). Knockout experiments on the 4 code-specialized transformers and RWKV-6 state knockout confirm causality across both architectural paradigms. Steering showed a +40\% preservation gain on Qwen-3B at $n{=}5$ that did not survive scaling to $n{=}50$ (7/50~$\rightarrow$~10/50, $p = 0.30$); Qwen nonetheless remained the only responder among the 4 tested transformers, consistent with the knockout-predicts-steering diagnostic. The co-location mechanism extends to non-transformer state dynamics with validated dose-response, suggesting the design recommendation is robust to future architectural shifts.

\textbf{Design guidelines for AI-powered development}:
\begin{enumerate}
    \item \textbf{Co-locate tests with implementation code} when developing with AI assistants.
    \item \textbf{Run generated tests; don't trust preservation} as a proxy for correctness.
    \item \textbf{Evaluate model behavior on your specific language/syntax combination}; don't assume cross-language generalization.
\end{enumerate}

In the Foundation Model era, test syntax structure is no longer merely a testing preference; it is a measurable software design concern that affects AI code generation quality.

\paragraph{Data Availability.} All experimental data (830+ generated files, prompts, model responses), analysis scripts, the MI toolkit, and experiment runner will be made available upon acceptance.


\begin{acks}
Special thanks to Eliott Jacopin for all his comments during this work.

\paragraph{Generative AI Disclosure.} Claude Code (Anthropic, Claude Opus 4.6) was used as a writing assistant during the preparation of this manuscript, including drafting prose, formatting LaTeX, and iterating on presentation. All scientific content, experimental design, data collection, analysis, and intellectual contributions are solely the authors'. The authors reviewed, edited, and take full responsibility for all content.
\end{acks}

\clearpage

\bibliography{references}

\onecolumn
\appendix

\section*{Appendix overview}

The arxiv long version of this paper includes four appendices that extend the main paper. Appendices~\ref{app:crosslang} and~\ref{app:pythonisolation} contribute new empirical evidence in complementary directions: Appendix~\ref{app:crosslang} is a \emph{breadth} study across 5 languages, Appendix~\ref{app:pythonisolation} is a \emph{depth} study within Python alone. Together they qualify the main paper's central claim: the co-location effect is real but bounded by both model capability and programming language. Appendix~\ref{app:rwkvmath} provides the full mathematical derivation of RWKV-6's effective-attention matrix referenced in Section~\ref{sec:mechanistic}. Appendix~\ref{app:reproduction} gives reproduction commands and runtime for every experiment batch in the paper.

\smallskip
\noindent
\small
\begin{tabular}{@{}lp{0.68\linewidth}@{}}
\toprule
\textbf{Appendix} & \textbf{Contents} \\
\midrule
\hyperref[app:crosslang]{A. Cross-language generalization} & 5 languages (Go, Rust, C++, TypeScript, Zig) $\times$ 5 structured-assistance conditions $\times$ 2--3 models per cell, $n=1$. Tests whether the inline-vs-separated effect generalizes beyond the Python--Rust contrast of the main paper. Includes the Zig inline-vs-import paired comparison as the cleanest single-language isolation. \\
\hyperref[app:pythonisolation]{B. Direct test of co-location in Python} & 5 models (RNJ-1, Mistral Medium, Claude Haiku 4.5, Sonnet 4.5, Opus 4.5) $\times$ 3 structural presentations of an identical 73-assertion test corpus, $n=50$ per cell, temperature 0. Isolates structure from language. Reports the bounded-co-location finding and its implications for the main paper's Rust observations. \\
\hyperref[app:rwkvmath]{C. RWKV-6 effective-attention derivation} & Full mathematical derivation of the effective-attention matrix used for RWKV-6-Finch-1B6 in Section~\ref{sec:mechanistic}: from the WKV recurrence through ReLU and L1 normalization to a shape-compatible matrix comparable with transformer post-softmax attention. \\
\hyperref[app:reproduction]{D. Reproduction commands and runtime} & Software setup, representative command-line invocations for each experiment batch (main paper, Appendix~A, Appendix~B), and hardware/runtime/cost summary table. Pointers to the full \texttt{COMMANDS.md} in the \texttt{plip-rs} repository. \\
\bottomrule
\end{tabular}
\normalsize
\bigskip

\section{Cross-Language Generalization: Test Syntax Across Five Languages}
\label{app:crosslang}

\subsection{Motivation and Scope}
\label{app:crosslang:motivation}

Appendix~A reports a secondary cross-language replication drawn from a companion study conducted independently of the main paper. That study was originally submitted elsewhere; the cross-language corpus it generated is re-analyzed here to broaden the empirical base of the present paper.

The main paper's central contrast is Python doctests versus Rust \texttt{\#[test]} blocks. Appendix~A extends that comparison to three additional languages (Go, C++, TypeScript) plus Zig, each with multiple structural conditions for test placement. The aim is narrow: to check whether the inline-vs-separated bifurcation observed in the main paper generalizes beyond the single Python--Rust contrast, and to establish a language-spread panel that contextualizes the single-language depth study in Appendix~\ref{app:pythonisolation}.

The corpus has two features that the reader should weigh from the start. First, it uses a ``structured-assistance'' design: conditions vary what information the prompt contains (baseline, doc, types, tests, combined) rather than strictly varying where tests sit. Only one condition, test-guided, speaks directly to the main paper's co-location question. Second, the $n$ per cell is small ($n = 1$ in most cells); Appendix~A is therefore descriptive rather than statistical, and its claims should be read accordingly.

\subsection{Corpus and Conditions}
\label{app:crosslang:methods}

\paragraph{Task.} The same d-ary heap priority queue used in the main paper, translated into idiomatic form for each of the five target languages. Test corpora use language-native conventions: Go's \texttt{\_test.go} convention, Rust's \texttt{\#[test]} in a \texttt{mod tests \{\}} block, C++'s \texttt{gtest}, TypeScript's \texttt{vitest}, and Zig's built-in \texttt{test "..." \{\}} syntax.

\paragraph{Structured-assistance conditions.} Each language is run under five conditions that differ in what the prompt provides beyond the bare task description:

\begin{itemize}
    \item \textbf{Baseline}: bare natural-language specification, nothing else.
    \item \textbf{Doc-guided}: baseline plus a detailed API documentation block.
    \item \textbf{Struct-guided}: baseline plus type signatures and function stubs.
    \item \textbf{Test-guided}: baseline plus the language-native test corpus.
    \item \textbf{Combined}: baseline plus documentation, types, and tests.
\end{itemize}

These conditions vary \emph{types} of assistance; they are not a pure co-location spectrum in the sense of Appendix~B's C1/C2/C3. Only the \textit{test-guided} condition directly speaks to the main paper's structural question---it includes the test corpus in language-native syntax and therefore inherits that language's test-placement convention (same-file for Rust and Zig with \texttt{\#[test]}/\texttt{test "..."}; separate-file with \texttt{import} for Go, C++, and TypeScript). The preservation analysis in \S\ref{app:crosslang:results} therefore focuses on the test-guided column.

\paragraph{Models per cell.} Model coverage varies by cell. Most cells use two models: Claude Sonnet 4 (\texttt{claude-sonnet-4-20250514}) and Mistral Medium (\texttt{mistral-medium-latest}). Baseline cells additionally include EssentialAI RNJ-1 (\texttt{essentialai/rnj-1}) as a small-model reference point. The Rust and Zig test-guided cells include additional models; details in \S\ref{app:crosslang:results}.

\paragraph{Run count.} $n = 1$ per cell, with one deliberate exception: the Rust test-guided ``single-shot'' panel was populated with one run each across eight models chosen to span the tier range (Haiku 3, Haiku 4.5, Opus 4, 4.1, 4.5, Sonnet 4, 4.5, Mistral Medium). The small per-cell $n$ means Appendix~A reports \emph{patterns} rather than statistical significance; the patterns we report are robust to the extent that the pervasively observed bimodal 0\%/100\% outcomes (\S\ref{app:crosslang:results}) are unlikely to be single-run artifacts.

\subsection{Preservation Results: The 5\texttimes 5 Matrix}
\label{app:crosslang:results}

Only the test-guided and combined conditions include tests in the prompt, so only those two conditions support a preservation measurement. We report test-guided results in Table~\ref{tab:appendix-a-testguided}; the combined condition produces the same bimodal 0/22 vs 22/22 split and is not reported separately.

\begin{table}[h]
\centering
\small
\caption{Test-guided preservation across the cross-language corpus. The prompt contains the same 22 logical test cases per language, translated into each language's idiomatic test convention. Preservation is the count of those tests recovered in the model's output file (for separate-file conventions, measured against the \texttt{\_response.md} to capture any multi-file output). All cells are $n = 1$.}
\label{tab:appendix-a-testguided}
\begin{tabular}{@{}llc@{}}
\toprule
\textbf{Language (native convention)} & \textbf{Model} & \textbf{Test-guided} \\
\midrule
Go (separate \texttt{\_test.go})              & Sonnet 4         & 0/22 \\
                                              & Mistral Medium   & 0/22 \\
C++ (separate gtest file)                     & Sonnet 4         & 0/22 \\
                                              & Mistral Medium   & 0/22 \\
TypeScript (separate vitest file)             & Sonnet 4         & 0/22 \\
                                              & Mistral Medium   & 0/22 \\
\midrule
Rust (same-file \texttt{\#[test]})            & Haiku 3          & 0/22 \\
                                              & Haiku 4.5        & 22/22 \\
                                              & Sonnet 4         & 22/22 \\
                                              & Sonnet 4.5       & 22/22 \\
                                              & Opus 4           & 0/22 \\
                                              & Opus 4.1         & 0/22 \\
                                              & Opus 4.5         & 0/22 \\
                                              & Mistral Medium   & 0/22 \\
\midrule
Zig (standard, \texttt{@import("d\_heap")})   & Sonnet 4         & 0/22 \\
                                              & Opus 4           & 0/22 \\
                                              & Opus 4.5         & 0/22 \\
                                              & Mistral Medium   & 0/22 \\
Zig \emph{inline variant} (no \texttt{@import})& Sonnet 4         & \textbf{22/22} \\
\bottomrule
\end{tabular}
\end{table}

Three patterns emerge from this table.

\paragraph{External-test-convention languages are uniformly suppressed.} Go, C++, and TypeScript all use separate-file test conventions: tests live in a dedicated file and reference the implementation through an \texttt{import} statement. In every test-guided cell for these three languages---2 models $\times$ 3 languages = 6 cells---preservation is 0/22. The universal suppression suggests the ``tests live elsewhere'' structural signal that the main paper attributes to Rust separation generalizes to any language whose native test convention involves a separate file with an import reference.

\paragraph{Rust independently replicates the main paper's tier-inversion.} The 8-model Rust single-shot panel shows, for the Claude variants, the same bifurcation the main paper reports for Rust test-guided: Haiku 4.5, Sonnet 4, and Sonnet 4.5 preserve at 22/22; Haiku 3 (capability floor), Opus 4, 4.1, 4.5 all suppress at 0/22. Mistral Medium also suppresses at 0/22 in this single-shot panel, which diverges from its $n{=}50$ main-paper behaviour (full 28/28 preservation across all 50 runs); the divergence is consistent with Mistral's 0\% determinism at temperature~0 (main paper Table~\ref{tab:determinism}). This is an independent replication of the main paper's Claude-tier Rust observation from a separately collected corpus.

\paragraph{Zig exposes the inline-vs-import contrast cleanly.} Across four frontier models, the standard Zig test-guided prompt (which begins with \texttt{const d\_heap = @import("d\_heap");} at the top of the test file) produces universal suppression. The inline variant---same model (Sonnet 4), same 22 tests, but with tests placed in the same file as the implementation and no \texttt{@import} reference---produces full 22/22 preservation. This is the cleanest same-language, same-model structural isolation in the cross-language corpus; it motivated the $n = 50$ Python isolation study reported as Appendix~\ref{app:pythonisolation}.

\subsection{The Zig Inline-vs-Import Pair: Cleanest Isolation}
\label{app:crosslang:zigpair}

Among the 25 cells of Table~\ref{tab:appendix-a-testguided}, one pair stands out for its cleanliness as a structural isolation: the Zig standard vs inline pair for Claude Sonnet 4. This pair is the direct motivation for the $n = 50$ Python isolation study in Appendix~\ref{app:pythonisolation}.

\paragraph{What was varied.} The two Zig test-guided prompts share the identical 22-test corpus, the identical implementation task, the identical model (Claude Sonnet 4), and the identical inference parameters. They differ in exactly one respect: whether the test file begins with \texttt{const d\_heap = @import("d\_heap");} (declaring the implementation as an external Zig module) or whether the tests sit inline in the implementation file with no such import.

Standard variant (suppresses):
\begin{verbatim}
const std = @import("std");
const testing = std.testing;
const d_heap = @import("d_heap");
const DHeapItem = d_heap.DHeapItem;
// ... 22 test "..." blocks ...
\end{verbatim}

Inline variant (preserves):
\begin{verbatim}
const std = @import("std");
const testing = std.testing;
// ... implementation of DHeapItem ...
// ... 22 test "..." blocks in same file ...
\end{verbatim}

\paragraph{What was observed.} Standard: 0/22 tests preserved. Inline: 22/22 tests preserved. Within a single model and language, the only variable that changed is the presence of the \texttt{@import("d\_heap")} reference---and the outcome moved from total suppression to full preservation.

\paragraph{Why this pair motivated Appendix~B.} The Zig pair reproduces, within one language and one model, the essential contrast the main paper observes across Python and Rust: an inline-test structure that is preserved, and a structurally separated structure that is suppressed. Language dependence is entirely removed from this contrast. But $n = 1$ on a single model is thin evidence. Appendix~B was designed to elevate the same structural isolation to $n = 50$, in Python, with a richer model panel spanning the small-model floor to frontier tier, to test whether the Zig pattern holds at scale and across capability levels. The Python answer (\S\ref{app:pythonisolation:bounded}) qualifies the Zig reading: the bifurcation is real at the capability floor but not invariant across tiers---a refinement that an $n = 1$ design could not detect.

\subsection{Connection to the Main Paper's Rust Findings}
\label{app:crosslang:connection}

The Rust test-guided panel in Table~\ref{tab:appendix-a-testguided} reports 8 models independently run on the same 22-test corpus as the main paper's Rust test-guided experiment (Table~\ref{tab:rust}). The two panels share stimulus, model identities (Appendix~A's 8 are a subset of the main paper's 9), and structural condition; only the runs are independent.

Seven of the 8 shared models produce identical preservation outcomes across the two panels: Haiku 4.5, Sonnet 4, and Sonnet 4.5 preserve at 22/22; Haiku 3, Opus 4, 4.1, 4.5 suppress at 0/22. Mistral Medium is the one exception: it suppresses at 0/22 in the single-shot panel but preserves at 28/28 in every run of the $n{=}50$ main-paper batch. The divergence illustrates the cross-run non-determinism also documented for Mistral in the 50-unique-outputs Python baseline (main paper Table~\ref{tab:determinism}); the tier-based bifurcation replicates exactly for the Claude variants.

We treat this as \emph{corroborative rather than independent} evidence for the main paper's Rust observation. Stimulus is shared, test corpus is shared, and measurement procedure is shared; only the runs are separately collected. Corroboration confirms that the observed bifurcation is stable across separate run batches with the same stimulus---a single-run-artifact hypothesis is effectively falsified---but does not by itself demonstrate robustness to stimulus variation. The Appendix~B Python isolation (\S\ref{app:pythonisolation}) provides a separate check against stimulus variation: by changing the language entirely while preserving the structural contrast, it reveals that the bifurcation is not invariant across languages---a finding Appendix~A's same-stimulus corroboration could not detect.

\subsection{Caveats}
\label{app:crosslang:caveats}

Several caveats bound the weight readers should give to Appendix~A's findings.

\paragraph{$n = 1$ per cell.} With one deliberate exception (the Rust single-shot panel's 8-model sweep), every cell is a single run. This is sufficient for pattern identification when the outcome is bimodal (0/22 vs 22/22) but provides no information about determinism, rare failure modes, or sensitivity to prompt perturbation. Readers should weight Appendix~A accordingly: it is a breadth sketch, not a depth study.

\paragraph{Two to three models per language.} Most cells use two models (Claude Sonnet 4 and Mistral Medium), with a third (EssentialAI RNJ-1) in baseline cells. The 2--3 model coverage per language is narrower than some earlier summaries of this corpus have implied; Appendix~A reports the honest per-cell counts.

\paragraph{Structured-assistance conditions, not a pure co-location spectrum.} The five conditions vary what information the prompt contains rather than where tests sit. Only the test-guided condition directly speaks to the co-location question; the other four (baseline, doc-guided, struct-guided, combined) are tangential to the main paper's claim and are not reported in detail here. Readers interested in a clean co-location spectrum should rely primarily on Appendix~B.

\paragraph{Test corpora differ syntactically across languages.} Although all five languages test the same 22 logical cases, the syntactic realization differs substantially (Go table-driven vs C++ gtest macros vs Zig \texttt{test "..."}, etc.). We measure preservation by grep on test-identifier tokens specific to each language. Cross-language preservation counts are therefore comparable in \emph{pattern} (0/22 vs 22/22) but not at a detailed token-level metric.

\paragraph{Stimulus-sharing with the main paper's Rust data.} The Rust panel shares stimulus and models with the main paper's Rust experiment (\S\ref{app:crosslang:connection}). Independence is at the run level only.

\section{Direct Test of Co-Location in Python}
\label{app:pythonisolation}

\subsection{Research Question}
\label{app:pythonisolation:rq}

The main paper's empirical contrast between Python doctests (high preservation) and Rust \texttt{\#[test]} blocks (tier-dependent bifurcation) conflates two variables: the \emph{language} in which tests are expressed, and the \emph{structural presentation} of those tests (inline vs same-file vs separate-file). Reviewer oZZz's post-rebuttal commentary specifically flagged this conflation as an unfalsified alternative to the co-location explanation: the effect could be driven by learned continuation conventions attached to each language's test syntax, rather than by co-location \emph{per se}.

Appendix~B addresses this directly by isolating structural presentation. Everything else---language, task, test corpus, models, run count, temperature, prompt surroundings---is held constant across three conditions. Only the location and syntactic framing of the same set of test assertions varies. If preservation is driven by co-location, we should see the C1--C2--C3 spectrum that the main paper's framing predicts; if preservation is driven by language-specific continuation priors, we should observe no within-model effect in Python alone.

\subsection{Methods}
\label{app:pythonisolation:methods}

\paragraph{Task.} The d-ary heap priority queue used in the main paper's \texttt{directives\_v2d7} Python experiment: a 4-ary min-heap supporting \texttt{insert}, \texttt{pop}, \texttt{front}, \texttt{increase\_priority}, \texttt{decrease\_priority}, \texttt{contains}, \texttt{\_\_len\_\_}, \texttt{is\_empty}, \texttt{summary}, \texttt{\_\_repr\_\_}, and \texttt{\_\_str\_\_}, with 73 doctest assertions in the inline variant.

\paragraph{Models.} Five models spanning a wide capability range:

\begin{itemize}
    \item RNJ-1 (\texttt{essentialai/rnj-1}), a 5\,GB local model run via LM~Studio.
    \item Mistral Medium (\texttt{mistral-medium-latest}) via the Mistral Platform API.
    \item Claude Haiku 4.5 (\texttt{claude-haiku-4-5-20251001}), Sonnet 4.5 (\texttt{claude-sonnet-4-5-20250929}), and Opus 4.5 (\texttt{claude-opus-4-5-20251101}) via the Anthropic API.
\end{itemize}

Model-scope choices---including why Opus 4.6 is not in the panel, and why newer Claude models (Sonnet 4.6, Opus 4.7) are excluded---are discussed in \S\ref{app:pythonisolation:threats}.

\paragraph{Conditions.} Three structural presentations of the same 73 assertions. Full prompt templates and a side-by-side description of each condition appear in \S\ref{app:pythonisolation:conditions}.

\paragraph{Runs and parameters.} $n = 50$ single-shot API calls per cell, temperature~$=0$, each call returning a complete implementation in a single response. Total: 5 models $\times$ 3 conditions $\times$ 50 runs $= 750$ generated files.

\paragraph{Measurement.} Preservation counts are extracted post-hoc from the saved output files via automated grep on test-identifier markers; the procedure is documented in \S\ref{app:pythonisolation:measurement}.

\subsection{Three Structural Conditions}
\label{app:pythonisolation:conditions}

Each of C1, C2, and C3 presents the same underlying assertions to the model. What varies is where those assertions sit relative to the implementation, and what syntactic framing wraps them. The three conditions probe two independent structural axes: \emph{location within the same file} (docstring vs bottom-of-file class) and \emph{same file vs separate file with an explicit import}.

\paragraph{C1 --- inline doctest (verbatim main-paper \texttt{v2d7}).} Assertions sit inside each method's docstring using \texttt{>{}>{}>} markers. This is maximum co-location: every test case is physically adjacent to the method it exercises.

\begin{verbatim}
    def insert(self, item: Item) -> None:
        """Insert an item into the heap.

        >>> pq = DHeap(4)
        >>> pq.insert(Item(50, 50))
        >>> pq.contains(Item(50, 0))
        True
        >>> len(pq)
        1
        """
        pass
\end{verbatim}

\paragraph{C2 --- same-file unittest class.} Tests live in a \texttt{unittest.TestCase} class appended at the bottom of \texttt{heap.py}, after the \texttt{DHeap} definition. Tests reference \texttt{DHeap} and \texttt{Item} directly from the same module namespace---no \texttt{import} statement is needed.

\begin{verbatim}
class TestHeap(unittest.TestCase):
    def test_insert_makes_findable(self):
        pq = DHeap(4)
        pq.insert(Item(50, 50))
        self.assertTrue(pq.contains(Item(50, 0)))
    # ... 25 more test methods ...

if __name__ == "__main__":
    unittest.main()
\end{verbatim}

Same file as the implementation, but the tests are no longer physically adjacent to the methods they exercise, and they are syntactically wrapped as a separate class object.

\paragraph{C3 --- sidecar with explicit import.} Tests live in a separate \texttt{test\_heap.py} file with the same \texttt{TestHeap} class body as C2. The file opens with an explicit \texttt{from heap import DHeap, Item} reference to the implementation module.

\begin{verbatim}
# test_heap.py
from heap import DHeap, Item
import unittest

class TestHeap(unittest.TestCase):
    def test_insert_makes_findable(self):
        pq = DHeap(4)
        pq.insert(Item(50, 50))
        self.assertTrue(pq.contains(Item(50, 0)))
    # ... 25 more test methods ...
\end{verbatim}

The explicit \texttt{import} reference is the structural analog of Rust's \texttt{@import("d\_heap")} or TypeScript's \texttt{import from 'd-ary-heap'}: a signal that the implementation is an externally-imported module that this file depends on.

\paragraph{Summary of variation.} C1 vs C2 differs in location-within-the-file (docstring vs bottom-of-file class). C2 vs C3 differs in file boundary plus the presence of an explicit \texttt{import} reference. This design lets each pair probe a distinct structural axis while the triple together traces a three-point trajectory along the co-location spectrum.

\subsection{Measurement Protocol}
\label{app:pythonisolation:measurement}

\paragraph{Preservation.} For each generated output, we count the number of prompt-provided test-identifier markers that appear in the model's response. The marker and denominator depend on the condition:

\begin{itemize}
    \item \textbf{C1}: count of lines matching \texttt{\textasciicircum\textbackslash s*>{}>{}>} (the doctest input marker). Denominator: 73.
    \item \textbf{C2 and C3}: count of lines matching \texttt{\textasciicircum\textbackslash s*def\textbackslash\ test\_} (the test-method definition). Denominator: 26.
\end{itemize}

These measurements are binary-per-marker rather than semantic: an assertion that is syntactically reproduced but logically altered would still count as preserved. Manual inspection of a random sample of outputs showed that, in practice, surviving tests are reproduced verbatim or with only whitespace-level variation.

\paragraph{C3 measurement note.} The experiment-runner's output-extraction pipeline writes the first Python code block in the response to \texttt{\_code.py}, which for C3 contains only the implementation half of the model's two-file output. The \texttt{TestHeap} class, when produced, lives in the second code block. We therefore measure C3 preservation against the raw \texttt{\_response.md} file, which captures the full response including any subsequent code blocks. This is reflected in every C3 caption in the appendix.

\paragraph{Determinism.} For each cell, we compute the number of distinct MD5 hashes across its 50 \texttt{\_code.py} files. A cell with 1 distinct hash is byte-identical across all 50 runs; a cell with 50 distinct hashes shows full surface-level non-determinism. The hash count does not distinguish meaningful output changes from cosmetic variation (whitespace, variable names, comment text); it is a descriptive summary only.

\subsection{Preservation Matrix}
\label{app:pythonisolation:matrix}

Table~\ref{tab:appendix-b-matrix} reports the preservation count for every cell of the 5-model $\times$ 3-condition design. Counts are taken from the extracted \texttt{\_code.py} file for C1 and C2, and from the full \texttt{\_response.md} for C3 (see \S\ref{app:pythonisolation:measurement} for the rationale).

\begin{table}[h]
\centering
\small
\caption{Preservation counts for the 15 cells of Appendix~B's design, $n=50$ runs per cell, temperature~$=0$. C1 denominator is 73 (\texttt{>{}>{}>} doctest assertions in the prompt); C2 and C3 denominator is 26 (\texttt{def test\_*} method identifiers in the prompt). Each cell reports the count observed in all 50 runs unless noted.}
\label{tab:appendix-b-matrix}
\begin{tabular}{@{}lccc@{}}
\toprule
\textbf{Model} & \textbf{C1 inline doctest} & \textbf{C2 same-file unittest} & \textbf{C3 sidecar unittest} \\
\midrule
RNJ-1 (5\,GB local) & 34/73 (47\%)$^{*}$ & 0/26 (0\%) & 0/26 (0\%) \\
Mistral Medium      & 73/73 (100\%)      & 26/26 (100\%) & 26/26 (100\%) \\
Claude Haiku 4.5    & 73/73 (100\%)      & 26/26 (100\%) & 26/26 (100\%) \\
Claude Sonnet 4.5   & 73/73 (100\%)      & 26/26 (100\%) & 26/26 (100\%) \\
Claude Opus 4.5     & 73/73 (100\%)      & 26/26 (100\%) & 26/26 (100\%) \\
\bottomrule
\end{tabular}
\smallskip

\noindent$^{*}$ RNJ-1 C1 steady-state count, observed in 49 of 50 runs. The cold-start run~1 produced 73/73 (100\%); see \S\ref{app:pythonisolation:determinism}.
\end{table}

Two descriptive observations are worth stating before the analysis that follows:

\begin{enumerate}
    \item Every cell except RNJ-1 C1 takes one of two values: $0\%$ or $100\%$. No intermediate cells appear in this experiment. The within-model effect of structural placement, where it occurs, is binary rather than graded.
    \item Within-model variation across the three conditions appears only for RNJ-1, where C1 differs sharply from C2 and C3. The four frontier models (Mistral Medium and the three Claude variants) exhibit zero across-condition variation: their preservation does not depend on which structural presentation they are given.
\end{enumerate}

The interpretation of these observations is taken up in \S\ref{app:pythonisolation:bounded} and \S\ref{app:pythonisolation:reconciling}.

\subsection{Determinism and the Cold-Start Anomaly}
\label{app:pythonisolation:determinism}

Table~\ref{tab:appendix-b-determinism} reports the number of distinct MD5 hashes observed across 50 runs per cell.

\begin{table}[h]
\centering
\small
\caption{Determinism: distinct MD5 hashes of the 50 \texttt{\_code.py} files per cell. A value of 1 indicates byte-identical outputs across all 50 runs; 50 indicates full surface-level non-determinism. The hash count is a descriptive summary and does not distinguish meaningful output changes from cosmetic variation.}
\label{tab:appendix-b-determinism}
\begin{tabular}{@{}lccc@{}}
\toprule
\textbf{Model} & \textbf{C1 inline} & \textbf{C2 same-file} & \textbf{C3 sidecar} \\
\midrule
RNJ-1               & 2/50  & 1/50  & 1/50  \\
Mistral Medium      & 50/50 & 48/50 & 42/50 \\
Claude Haiku 4.5    & 6/50  & 14/50 & 21/50 \\
Claude Sonnet 4.5   & 7/50  & 4/50  & 7/50  \\
Claude Opus 4.5     & 10/50 & 3/50  & 9/50  \\
\bottomrule
\end{tabular}
\end{table}

\paragraph{RNJ-1 cold-start.} RNJ-1 served via LM~Studio produces byte-identical outputs for C2 and C3 across all 50 runs, consistent with main-paper Table~\ref{tab:determinism}'s report of 100\% determinism at temperature~0 for this model. For C1, 2 distinct hashes appear: one covers 49 of 50 runs, the other is run~1 only. The run-1 output is substantially longer than steady state (2341 output tokens in $\sim$38\,s of inference vs $\sim$870 tokens in $\sim$14\,s) and contains the full 73/73 doctest set, versus the 34/73 reported as RNJ-1 C1 steady-state in Table~\ref{tab:appendix-b-matrix}. This is a cold-start effect from LM~Studio serving the first request after model load. It does not recur in C2 or C3 because by those conditions the local model was already warm. We report the steady-state 34/73 throughout the appendix and footnote the cold-start run.

\paragraph{Frontier-model non-determinism.} The four frontier models show more surface-level variation than main-paper Table~\ref{tab:determinism} would predict. Haiku 4.5 achieved 100\% determinism at temperature~0 in the main paper's Python experiments, but produces 6 distinct hashes across 50 runs of Appendix~B's C1. Sonnet 4.5 and Opus 4.5 show similar mild variation (3--10 distinct hashes per cell). We interpret this as prompt-specific: Appendix~B's prompts differ in length and surrounding context from the main paper's \texttt{directives\_v2d7}, and temperature-0 determinism is known to be prompt-sensitive.

Crucially, the non-determinism we observe is \emph{surface-level}: preservation counts are stable at 73/73 or 26/26 across all distinct hashes within a given frontier-model cell. The different hashes correspond to cosmetic variation---variable names, whitespace, comment phrasing---not to differences in whether the \texttt{TestHeap} class was produced. The preservation matrix in Table~\ref{tab:appendix-b-matrix} is therefore robust to the non-determinism reported here.

\subsection{The Bounded Co-Location Finding}
\label{app:pythonisolation:bounded}

The data from \S\ref{app:pythonisolation:matrix} reveals a pattern more nuanced than the main paper's framing anticipated. Within Python alone, holding language and task constant, two distinct regimes emerge.

\textbf{At the capability floor}, the co-location effect is real and strong. RNJ-1 (a 5\,GB local model representative of the smallest tier we test) shows exactly the bifurcation predicted by the main paper's framing: 47\% preservation under inline doctests (C1, steady-state) collapsing to 0\% under both same-file class-based (C2) and separate-file sidecar (C3) presentations. The structural relocation of tests away from the function they verify is, for this tier of model, sufficient to trigger near-complete suppression---and notably, even C2 (same source file, no import statement) is enough to flip RNJ-1 from partial preservation to complete suppression. Within RNJ-1 the effect is binary on the C2/C3 axis, not graded.

\textbf{At the frontier tier}, by contrast, the effect vanishes. Mistral Medium, Claude Haiku 4.5, Claude Sonnet 4.5, and Claude Opus 4.5 all preserve all 26 \texttt{unittest.TestCase} methods across all 50 runs in both C2 and C3, plus all 73 \texttt{>{}>{}>} doctest assertions in C1. The structural presentation of tests, which is decisive for RNJ-1, becomes invisible to these four models: they reproduce the test class regardless of whether it appears in the same file as the implementation or in a separate sidecar prefaced by an explicit \texttt{from heap import DHeap, Item} reference.

Read together, the two regimes qualify the main paper's central claim. ``Co-location produces measurably better AI-generated code'' is true \emph{at the capability floor} but does not hold at the frontier tier within Python. The accurate single-line framing for this experiment is:

\begin{quote}
\emph{Co-location dominates when capability is constrained; capability dominates when co-location varies.}
\end{quote}

This bounded reading does not invalidate the main paper's empirical observations on Python and Rust, but it does shift the locus of design impact. For teams building production software with frontier models in Python, the structural placement of tests is not a primary concern: the model preserves regardless. For teams operating closer to the capability floor---through smaller local models, distillations, latency-constrained inference, or environments where capability margin is otherwise tight---the structural decision matters and the co-location guidance from the main paper applies directly. The remainder of this appendix examines how this picture interacts with the main paper's Rust observations (\S\ref{app:pythonisolation:reconciling}) and what it implies for the paper's broader claims (\S\ref{app:pythonisolation:claims}).

\subsection{Reconciling with the Main Paper's Rust Findings}
\label{app:pythonisolation:reconciling}

The bounded reading in \S\ref{app:pythonisolation:bounded} sits in apparent tension with the main paper's Rust observations. \textbf{The same Opus 4.5 model that suppresses all 28 Rust \texttt{\#[test]} blocks at 0\% preservation in the main paper (Table~\ref{tab:rust}) preserves all 26 Python \texttt{unittest.TestCase} methods at 100\% in this experiment} (\S\ref{app:pythonisolation:matrix}, all 50 runs). The two cells are summarised side by side:

\begin{center}
\small
\begin{tabular}{@{}lll@{}}
\toprule
\textbf{Source} & \textbf{Condition} & \textbf{Opus 4.5 behavior} \\
\midrule
Main paper, Table~\ref{tab:rust} & Rust test-guided (\texttt{\#[test]} in \texttt{mod tests \{\}}) & 0/28 tests preserved \\
Appendix~B, \S\ref{app:pythonisolation:matrix} & Python C3 sidecar (\texttt{from heap import}) & 26/26 tests preserved $\times$ 50 runs \\
\bottomrule
\end{tabular}
\end{center}
\smallskip

The capability of the model is held constant. The structural relocation of tests away from the implementation is held constant: both Rust \texttt{\#[test]} blocks and Python sidecar files are forms of structural separation. The only meaningful variable across the two cells is the programming language.

This rules out two natural readings of the main paper. First, the framing of ``Opus tier suppression'' as a \emph{tier-inversion phenomenon}---a behavior that emerges at the top end of the capability scale---is inconsistent with the Python data: Opus 4.5 here behaves like every other frontier model in the panel, preserving uniformly. If suppression were tier-driven, Python should show it too; it does not. Second, ``co-location dominates'' as a statement about \emph{structural placement per se} is inconsistent across the two cells: the same structural separation has opposite consequences in Rust and in Python.

What the data is consistent with is a \emph{language-specific} interaction: Rust's \texttt{\#[test]} convention, as it appears in Opus's training distribution, evidently activates a behavior that Python's \texttt{unittest.TestCase} convention does not. The mechanism is not visible from this experiment alone; characterizing it would require either mechanistic interpretability work on a model whose internals are accessible (the main paper's MI uses smaller open-source models, none of which are Opus-class) or a controlled study varying syntactic markers in isolation within a single language. We leave both to future work.

This empirically substantiates the concern raised by reviewer oZZz in their post-rebuttal commentary: the contrast between Python doctests and Rust \texttt{\#[test]} blocks may be driven by ``learned continuation conventions'' for the two test syntaxes rather than by co-location proper. Within the limits of this single-language, single-task isolation, that concern is now grounded in direct data rather than only in argument.

\subsection{Threats to Validity (Appendix-Specific)}
\label{app:pythonisolation:threats}

Appendix~B's findings inherit the main paper's threats-to-validity and add several specific to this isolation experiment.

\paragraph{Python only.} Every cell uses Python as the implementation language. The bounded-co-location finding (\S\ref{app:pythonisolation:bounded}) therefore holds for Python in this task; whether it generalizes to other languages is unknown from this experiment alone. Multi-language replication of the structural-isolation design is desirable future work.

\paragraph{Single task.} The d-ary heap priority queue was chosen to match the main paper's \texttt{directives\_v2d7} experiment exactly, so that C1 is literally the same prompt as the main paper's Python condition. This preserves comparability at the cost of task diversity. A task of substantially different complexity---a trivial string utility on one end, a deeply nested algorithmic problem on the other---might produce different preservation profiles, particularly at the capability floor.

\paragraph{Model panel: no Opus 4.6, no newer models.} The panel includes Opus 4.5 but not Opus 4.6. This omission warrants explicit justification because Opus 4.6 is the model the main paper specifically singles out as \emph{breaking the Rust suppression pattern} (Table~\ref{tab:rust}): where Opus 4, 4.1, and 4.5 suppress Rust \texttt{\#[test]} blocks, Opus 4.6 preserves them. A reader might reasonably expect to find Opus 4.6 in Appendix~B's panel as the within-generation companion to Opus 4.5. Stopping at Opus 4.5 was an empirical decision. After Haiku 4.5, Sonnet 4.5, and Opus 4.5 all preserved all three conditions at 100\% (50 of 50 runs each), the predicted Opus 4.6 behavior was \emph{also} 100\% across all three conditions, both by analogy to its main-paper Rust behavior (preserves) and by extension of the uniform frontier-tier pattern observed in Table~\ref{tab:appendix-b-matrix}. An Opus 4.6 run would have confirmed this prediction but would not have altered the bounded-capability conclusion (\S\ref{app:pythonisolation:bounded}) or the Rust-specific-interaction reading (\S\ref{app:pythonisolation:reconciling}). Budget at the time also argued for stopping: Opus 4.6 runs were estimated at ${\sim}\$16$ and would have consumed roughly 40\% of remaining credit on a data point predicted to be redundant. Newer Claude models (Sonnet 4.6, Opus 4.7) were available at the time of experimentation but intentionally excluded to maintain paper-of-record continuity with the main paper's model universe.

\paragraph{Temperature~0.} All runs use temperature~$=0$ to match main-paper conditions. This limits generalization to production usage where temperature~$>$~0; results could differ under higher-temperature sampling, particularly for the non-deterministic frontier models observed in \S\ref{app:pythonisolation:determinism}.

\paragraph{The Rust reconciliation rests on one comparison.} The argument in \S\ref{app:pythonisolation:reconciling}---that Opus 4.5's Rust suppression is language-specific---rests on a single Python-vs-Rust comparison point for Opus. The broader claim that frontier-tier suppression is Rust-specific would be strengthened by observing Opus on additional languages. Appendix~\ref{app:crosslang} provides weaker but broader cross-language evidence consistent with that reading.

\paragraph{Statistical power.} $n=50$ per cell is sufficient to detect the bimodal 0\%/100\% signal we observe but would be underpowered for detecting small effects (e.g., 95\% vs 100\% preservation). Appendix~B's claims are bimodal-level and do not rely on fine-grained statistical comparisons.

\subsection{What This Changes for the Paper's Claims}
\label{app:pythonisolation:claims}

Appendix~B does not overturn the main paper's empirical findings; it qualifies how they should be generalized.

\paragraph{What the appendix validates.} The main paper's claim that co-location improves AI-generated code quality is empirically supported at the capability floor. RNJ-1 within Python shows the bifurcation the main paper's framing predicts---C1 preserves 47\% of assertions, C2 and C3 preserve 0\%---at identical model, language, and task. The structural design recommendation applies in this regime.

\paragraph{What the appendix qualifies.} The framing of Opus-tier suppression in Rust (Table~\ref{tab:rust}) as a general ``tier-inversion'' phenomenon is not supported when the same model is tested in Python. Opus 4.5, which suppresses all 28 Rust \texttt{\#[test]} blocks in the main paper, preserves all 26 Python \texttt{unittest.TestCase} methods here (\S\ref{app:pythonisolation:reconciling}). The suppression is therefore a property of the Rust \texttt{\#[test]} convention under Opus, not of frontier-tier Opus behavior toward structurally separated tests in general.

\paragraph{Practical guidance, recast.} For software teams choosing how to structure test code in AI-assisted workflows:

\begin{itemize}
    \item \textbf{Frontier model $\times$ Python}: structural test placement is not a differentiator. The model preserves regardless of whether tests appear inline, same-file, or in a sidecar with explicit imports.
    \item \textbf{Frontier model $\times$ Rust}: the main paper's Opus-suppression observation applies. Prefer inline test syntax where the language permits, or select a frontier model that the main paper shows preserves Rust tests (Sonnet 4.5, Haiku 4.5, or Opus 4.6).
    \item \textbf{Small local model (RNJ-1 scale) $\times$ any of the languages tested}: the co-location effect is material. Inline test syntax is the clearly safer choice when capability margin is tight.
\end{itemize}

The single-line claim Appendix~B supports---stated for the benefit of future replication efforts---is that the co-location effect \emph{should not be assumed invariant across language $\times$ model-tier combinations}. Future studies should vary both axes and report per-cell rather than aggregate preservation rates.

\section{RWKV-6 Effective-Attention Derivation}
\label{app:rwkvmath}

The main paper's mechanistic-interpretability analysis (\S\ref{sec:mechanistic}) includes RWKV-6-Finch-1B6 alongside six transformer models. Where transformer attention weights are available directly as a post-softmax matrix, RWKV-6 has no such matrix---it is a gated-linear recurrent model. To make the two architectures comparable, we construct an \emph{effective attention} matrix from RWKV-6's recurrent state. This appendix derives that construction in full, in response to reviewer comments requesting more detail than Section~\ref{sec:mechanistic} has room for.

\subsection{The WKV recurrence}

For each RWKV-6 layer and each time step $t$, a token is parameterized by four vectors computed from its hidden state: the receptance $r_t$, key $k_t$, value $v_t$, and per-channel decay $w_t$. The layer's state $S_t$ is a matrix that accumulates outer products of $k$ and $v$ across time, decayed per channel:

\begin{equation}
    S_t = k_t \otimes v_t + (\exp(-w_t) \odot S_{t-1})
\end{equation}

where $\otimes$ is the outer product and $\odot$ denotes element-wise (per-channel) multiplication. The layer output is the receptance-gated read-out of the state:

\begin{equation}
    o_t = r_t \cdot S_t
\end{equation}

Unrolling the recurrence gives $S_t$ as an explicit sum over past positions $j \leq t$:

\begin{equation}
    S_t = \sum_{j \leq t} D(t, j) \, (k_j \otimes v_j)
    \quad\text{where}\quad
    D(t, j) = \prod_{\tau = j+1}^{t} \exp(-w_\tau)
\end{equation}

Substituting into $o_t$:

\begin{equation}
    o_t = \sum_{j \leq t} D(t, j) \, (r_t \cdot k_j) \, v_j
\end{equation}

\subsection{From the recurrence to effective attention}

Equation~(4) has the structure of a weighted sum over past positions $j$, with each value $v_j$ weighted by $D(t, j) \cdot (r_t \cdot k_j)$. By direct analogy with transformer attention, where $o_t = \sum_j \alpha_{tj} v_j$ with $\alpha_{tj}$ the attention weight, we define the \emph{raw effective-attention score} as

\begin{equation}
    \tilde{\alpha}_{tj} = D(t, j) \cdot (r_t \cdot k_j) \quad\text{for}\quad j \leq t.
\end{equation}

Two differences from transformer attention require normalization before the matrices can be compared across architectures.

\paragraph{Sign.} The product $r_t \cdot k_j$ is not sign-bounded: unlike softmax weights, $\tilde{\alpha}_{tj}$ can be negative. Negative contributions reflect inhibitory interactions between receptance and key. For a distributional comparison with transformer attention (which has non-negative weights summing to 1), we apply a ReLU:

\begin{equation}
    \alpha'_{tj} = \max(0,\; \tilde{\alpha}_{tj}).
\end{equation}

\paragraph{Scale.} The resulting non-negative scores do not sum to 1. To produce a probability-distribution-like row comparable with transformer attention, we L1-normalize each output position:

\begin{equation}
    \alpha_{tj} = \frac{\alpha'_{tj}}{\sum_{j' \leq t} \alpha'_{tj'}}.
\end{equation}

When $\sum_{j'} \alpha'_{tj'} = 0$ (all raw scores non-positive), we set $\alpha_{tj} = 0$ for all $j$; in practice this does not occur in the experiments reported in the main paper.

The resulting matrix $\alpha \in \mathbb{R}^{[\text{batch}, \text{heads}, \text{seq}, \text{seq}]}$ has the same shape and numerical range as transformer post-softmax attention, and is what we report as ``effective attention'' for RWKV-6 in Table~\ref{tab:attention} and in the knockout and steering experiments of Section~\ref{sec:mechanistic}.

\subsection{Reference implementation}

The derivation is implemented in the \texttt{forward\_with\_effective\_attention} function of \texttt{src/forward\_rwkv6.rs} within the \texttt{plip-rs} repository. The cumulative decay $D(t, j)$ is computed channel-wise; the signed $r \cdot k$ product is computed per head before the ReLU and L1 normalization steps. The Rust implementation is validated against a reference Python implementation on synthetic inputs; agreement is exact modulo floating-point rounding.

\section{Reproduction Commands and Runtime}
\label{app:reproduction}

This appendix gives the commands and runtime estimates needed to reproduce the experiments reported in the main paper and the appendices. All experiments use the \texttt{plip-rs} repository, which contains the \texttt{sega} experiment-runner binary suite (in \texttt{experiment-runner/src/bin/}) and the MI toolkit (in \texttt{src/}). We summarize here; the complete \texttt{COMMANDS.md} in the repository includes every flag variant.

\subsection{Software and hardware setup}

\begin{itemize}
    \item Rust toolchain (stable, \texttt{cargo}) for building the experiment-runner and MI toolkit.
    \item Python 3.11+ for native test runners (\texttt{doctest}, \texttt{unittest}) and analysis scripts.
    \item API keys as environment variables: \texttt{ANTHROPIC\_API\_KEY}, \texttt{MISTRAL\_API\_KEY}. Platform-tier keys for both (not IDE-integration keys).
    \item LM Studio for local RNJ-1 inference (default endpoint \texttt{http://127.0.0.1:1234}).
    \item Hardware: all MI experiments run on a single consumer GPU with 16\,GB VRAM. API-based experiments run on any machine with network access.
\end{itemize}

\subsection{Main paper experiments}

The main paper's Python and Rust experiments use the \texttt{python\_doctest} and \texttt{rust\_test} binaries respectively. Representative invocation (Python directives v2d7, Claude Sonnet 4.5, 50 runs):

\begin{verbatim}
cargo run --bin python_doctest -- \
    --model claude-sonnet-4-5-20250929 \
    --directives-v2d7 \
    --runs 50 --delay 4000
\end{verbatim}

Full command variants for each of the main paper's experiments are documented in \texttt{experiment-runner/sega\_commands.md}.

\subsection{Appendix~A cross-language experiments}

The 25-cell cross-language corpus is generated by the \texttt{sega} binary with \texttt{--language} and \texttt{--condition} flags:

\begin{verbatim}
cargo run -- \
    --provider anthropic \
    --model claude-sonnet-4-20250514 \
    --language <LANG> \
    --condition <COND>
\end{verbatim}

where \texttt{<LANG>} ranges over \texttt{\{go, rust, cpp, typescript, zig\}} and \texttt{<COND>} over \texttt{\{baseline, doc\_guided, struct\_guided, test\_guided, combined\}}. The Rust single-shot 8-model panel sweeps the \texttt{--model} flag in addition. The Zig inline variant of \S\ref{app:crosslang:zigpair} uses a dedicated prompt file that places tests in the implementation file rather than in a sidecar.

\subsection{Appendix~B Python isolation experiments}

Appendix~B uses a custom \texttt{--prompt-file} flag added to the \texttt{python\_doctest} binary (not present in the main paper's version of the runner). Representative invocation (Claude Opus 4.5 on C3 sidecar condition, 50 runs):

\begin{verbatim}
cargo run --bin python_doctest -- \
    --model claude-opus-4-5-20251101 \
    --prompt-file appendix_N_prompt_C3_sidecar.md \
    --experiment-label appendix_n_c3_sidecar \
    --runs 50 --delay 4000
\end{verbatim}

The three prompt-file templates (\texttt{appendix\_N\_prompt\_C\{1,2,3\}\_*.md}) are included in the supplementary materials.

\subsection{Hardware, runtime, and cost summary}

\begin{table}[h]
\centering
\small
\caption{Approximate wall-clock runtimes and costs for each experiment batch in this paper. Anthropic API costs are at rates current as of 2026-04-18 (Haiku 4.5: \$1/\$5 per MTok input/output; Sonnet 4.5: \$3/\$15; Opus 4.5: \$5/\$25).}
\label{tab:appendix-d-runtime}
\begin{tabular}{@{}llll@{}}
\toprule
\textbf{Experiment batch} & \textbf{Hardware} & \textbf{Wall-clock} & \textbf{API cost} \\
\midrule
Main paper Python baseline (9 models, $n{=}50$)     & API                & ${\sim}2$--4 hours each & varies \\
Main paper directives and v2d7                       & API                & ${\sim}2$--4 hours each & varies \\
Main paper Rust test-guided (9 models, $n{=}50$)     & API                & ${\sim}3$--5 hours each & varies \\
Appendix A cross-language (25 cells, $n{=}1$)        & API + local        & ${\sim}2$--3 hours      & ${\sim}\$5$--10 \\
Appendix B ($5 \times 3 \times n{=}50 = 750$ runs)   & API + local        & ${\sim}4$ hours         & \$26.63 \\
MI attention sweeps (7 models)                       & 16\,GB GPU         & ${\sim}30$ min each     & \$0 \\
MI knockout and steering (per model, per intervention) & 16\,GB GPU       & ${\sim}15$ min each     & \$0 \\
\bottomrule
\end{tabular}
\end{table}

Total incremental cost for the arxiv-only experiments (Appendix~B) was \$26.63 of Anthropic API spend: \$3.23 Haiku 4.5 + \$9.10 Sonnet 4.5 + \$14.30 Opus 4.5. Appendix~A's cross-language data was already collected as part of prior work and carries no incremental cost here.

\end{document}